\documentclass{aa}  
\usepackage{graphicx}
\usepackage{txfonts}
\usepackage{natbib}
\bibpunct[; ]{(}{)}{,}{a}{}{,}
\def\asec{\ifmmode ^{\prime\prime}\else$^{\prime\prime}$\fi}

\def\msun{M$_{\odot}$}

\def\grad{$^\circ$}
\def\it{\sl}
\def\degs{\ifmmode ^{\circ}\else$^{\circ}$\fi}
\def\amin{\ifmmode ^{\prime}\else$^{\prime}$\fi}
\def\asec{\ifmmode ^{\prime\prime}\else$^{\prime\prime}$\fi}
\def\fm{\hbox{$.\!\!^{\rm m}$}}            
\def\fdg{\hbox{$.\!\!^\circ$}}          
\def\farcs{\hbox{$.\!\!^{\prime\prime}$}}  

\def\psr{PSR~J1357$-$6429}
\def\degs{\ifmmode ^{\circ}\else$^{\circ}$\fi}
\def\amin{\ifmmode ^{\prime}\else$^{\prime}$\fi}
\def\farcm{\hbox{$.\mkern-4mu^\prime$}}
\def\eqalign#1{\null\,\vcenter{\openup1\jot \m@th
   \ialign{\strut\hfil$\displaystyle{##}$&$\displaystyle{{}##}$\hfil
   \crcr#1\crcr}}\,}
\begin{document}
   \title{Possible optical counterpart      
   of PSR J1357$-$6429
\thanks{Based 
on observations made with ESO telescope at the Paranal Observatory under
Programme 083.D-0449A and with archival ESO VLT data, obtained from
the ESO/ST-ECF Science Archive Facility.}}

\author{A.~Danilenko\inst{1} 
\and A.~Kirichenko\inst{1,2}
\and R.~E.~Mennickent\inst{3}
 \and G.~Pavlov\inst{2,4}
\and Yu.~Shibanov\inst{1,2}
\and S.~Zharikov\inst{5}
\and D.~Zyuzin\inst{1,2}
}

\offprints{A. Danilenko, \\  \email{ danila@astro.ioffe.ru}}

 \institute{ 
Ioffe Physical Technical Institute, Politekhnicheskaya 26,
St. Petersburg, 194021, Russia \\
danila@astro.ioffe.ru   
\and
St. Petersburg State Polytechnical Univ., Politekhnicheskaya 29, 
St. Petersburg, 195251, Russia \\ 
aida.taylor@gmail.com, dima$\_$zyuz@mail.ru  
\and
Department of Astronomy, Universidad de Concepcion, Casilla 160-C, Concepcion, Chile \\
rmennick@astro-udec.cl 
\and Department of Astronomy \& Astrophysics, Pennsylvania State University, PA 16802, USA \\ 
pavlov@astro.psu.edu
\and 
Observatorio Astron\'{o}mico Nacional SPM, Instituto de Astronom\'{i}a, Universidad Nacional 
Aut\'{o}nomia de Mexico, Ensenada, BC, Mexico \\ 
zhar@astrosen.unam.mx
}


 
  \abstract
   {PSR J1357$-$6429 is a Vela-like radio pulsar that has been recently detected in X-rays and $\gamma$-rays. 
   It powers a compact  tail-like X-ray pulsar wind nebula and  X-ray-radio plerion associated with 
   an extended TeV source HESS J1356$-$645. }
   {We present our deep optical 
 observations  with the Very Large Telescope 
to search for an optical counterpart of the pulsar and its nebula.}
   {The observations were carried out using a direct imaging mode 
 in the $V$, $R$, and $I$ bands. We also analysed archival  X-ray data  
obtained with \textit{Chandra} and \textit{XMM-Newton}.  
}
   {In all three optical bands, 
we detect  a   point-like source 
with $V$ = 27.3 $\pm$ 0.3, $R$ = 25.52 $\pm$ 0.07, and $I$ = 24.13 $\pm$ 0.05, 
whose position is    
within the 1$\sigma$ error circle of 
the X-ray position of the pulsar,
and  whose   colours are distinct from those of ordinary stars.   
We consider it as a candidate 
optical counterpart of the pulsar. If it is 
indeed   the    counterpart, its  5$\sigma$  offset 
from  the radio  pulsar position, measured  about 9 yr earlier, implies that the transverse 
velocity of the pulsar is 
in the range of 1600--2000 km s$^{-1}$  at the distance of 2--2.5 kpc, making it the fastest moving pulsar known. 
The direction of the estimated proper motion coincides with the extension of the
pulsar's   X-ray  tail,  suggesting that this is a jet. 
The tentative optical
luminosity and efficiency  of the pulsar are similar to those of the Vela pulsar,  
which  also supports  the  
optical identification.    
However,  the  candidate undergoes an unusually steep dereddened flux increase towards the infrared 
with a spectral index  $\alpha_{\nu}$ $\sim$ 5, that is   
not typical of optical pulsars.  It 
implies  a strong double-knee spectral break in the pulsar emission between 
the optical and X-rays.   
The reasons for the spectral steepness are unclear. 
It   may be caused by a nebula knot projected onto the jet and strongly overlapping with the pulsar, 
as observed for the Crab, where the knot has a significantly steeper spectrum than the pulsar. 
We find no other signs of the pulsar nebula in the optical. Alternatively, 
the detected source may be a faint AGN, 
that has not yet been seen at other wavelengths.    
}  
%
{The  position  and 
peculiar  colours 
  of the  detected source  
 suggest that  it is  an optical counterpart of the pulsar. 
  Further high spatial-resolution 
 infrared observations can help to verify its real nature.  
}


\authorrunning{A. Danilenko, A. Kirichenko, R.~E.~Mennickent, et al.}
\titlerunning{ Possible optical  counterpart }
   \maketitle

%
\section{Introduction}
\label{sec1}

PSR J1357$-$6429 is a young (characteristic age $\tau$ = 7.3 kyr) and energetic (spin-down luminosity 
$\dot{E}$ = 3.1 $\times$ 10$^{36}$ ergs~s$^{-1}$) 166 ms  radio pulsar that was 
discovered in the Parkes multi-beam survey 
of the Galactic plane \citep{Camilo04}. At a distance of 2.4 kpc estimated from 
its dispersion measure (DM), 
it is one of the nearest young pulsars known. This proximity has motivated   further  observations   
of the pulsar field in different spectral domains.  
   
The first X-ray observations with \textit{Chandra}  \citep{Zavlin07}  and \textit{XMM-Newton}  \citep{Esposito07}
have revealed    an X-ray counterpart of the pulsar  and    a  faint tail-like  signature of the   
pulsar wind nebula (PWN)   extended by a few arcseconds  northeast of the pulsar.  
Deeper X-ray observations \citep{Chang2011,Lemoine-Goumard11} have firmly established 
X-ray pulsations with the pulsar period 
and  found the tail-like structure 
to be  extended  out to several tens  arcseconds from the pulsar.
In addition,  a fainter extended X-ray emission  was detected 
on a few tens of arcminutes scale  \citep{Chang2011,Abramowski11},  
showing  a  plerion-like structure, which is   typical of many young pulsars.   
Finally,  periodic pulsations of  PSR J1357$-$6429 were discovered in the GeV range with \textit{Fermi} 
\citep{Lemoine-Goumard11}, and an extended source, \object{HESS J1356$-$645},   associated with the pulsar 
has been found in the TeV range with H.E.S.S. \citep{Abramowski11}. The HESS source 
positionally
coincides with the X-ray plerion, 
whose   radio counterpart  has also been found  in archival data. The association with  a  supernova remnant
(SNR)  candidate,  catalogued as  
\object{G309.8$-$2.6}, is debated.  
The age and observational properties of  the  J1357$-$6429 pulsar/PWN system 
appear to be  similar to the Vela pulsar system, 
which is about ten times closer to us and much more comprehensively 
studied in various spectral domains.
Further studies of the J1357$-$6429 system 
are important to improve our understanding the physics of young  pulsars and their PWNe.   
     
In contrast to Vela,  the J1357$-$6429  field has  not been studied in the optical. 
To search  for  an optical counterpart  of  the  J1357$-$6429 pulsar/PWN system, we carried out  
the first deep observations of its field with the ESO Very Large Telescope (VLT) in $VRI$ bands. 
Using some 
of our data available from the VLT archive, 
\citet{Mignani11} reported their failure to detect a counterpart in these observations.  
Analysing our complete data  set, we find  that 
this was a hasty conclusion.  We  firmly detected  a point-like 
source,  whose  position is 
within the 1$\sigma$ error circle of
the X-ray position of the pulsar. 
Its   colours  are distinct from those of field stars. 
We compare the optical data with the X-ray  data retrieved from the \textit{Chandra} and 
\textit{XMM-Newton} archives\footnote{\textit{Chandra}: ACIS-I, Obs 10880,  60 ks exposure, PI G. Pavlov; HRC-S, 
OBs 6656 and 7219, exposures 17 and 16 ks, 
PI M. Mendez. \textit{XMM}:  EPIC-MOS and PN, Obs 0603280101, exposure 78 ks and 55 ks, PI  G. Pavlov}.  
The observations and  data reduction are described  
in Sect.~\ref{sec2}, our results are presented in Sect.~\ref{sec3}, 
and discussed  in  Sect.~\ref{sec4}.
\begin{table}[t]
\caption{Log of the VLT/FORS2 observations of 
\object{\psr}. }
\begin{center}
\begin{tabular}{llcll}
\hline\hline
   Date           &  Band            & Exposure       &  Mean          & Seeing          \\
                  &                  &                &  airmass       & range           \\
                  &                  & [s]            &                & [arcsec]      \\
\hline \hline                
   2009-04-04     &  $V$           & 579$\times$5            & 1.35           &  0.5--0.7       \\
                  &  $R$           & 579$\times$5            & 1.43           &  0.5--0.6       \\
   2009-04-22     &  $V$          & 589$\times$20            & 1.35           &  0.5--0.8       \\                              
   2009-04-24     &  $R$          & 589$\times$10           & 1.35           &  0.5--0.7       \\
   2009-04-25     &  $I$          & 199$\times$39           & 1.32           &  0.4--0.6       \\ 
\hline
\end{tabular}
\end{center}
\label{t:log}
\end{table}
\begin{figure*}[t]
\begin{center}
\includegraphics[width=160mm,  clip=]{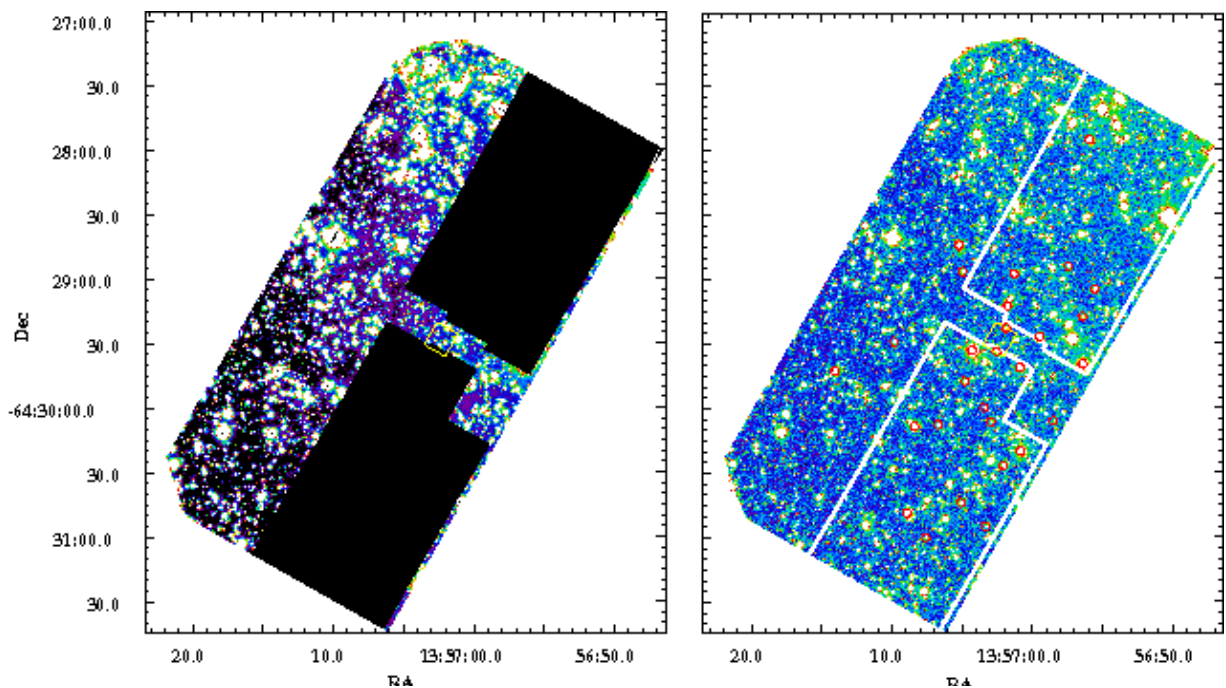} 
 \end{center}
 \caption{2\farcm$1\times$ 4\farcm2~ VLT/FORS2 $I$-band  images of the  PSR J1357$-$6429    
 field obtained with  long ({\sl left}) and short ({\sl right}) integration times. Only the FORS2/CCD1 chip    
 is shown.  The pulsar is located at the centre of a small yellow box. 
 Dark rectangular regions on the left  
 image are covered with  occulting bars to eliminate the contamination in the pulsar neighbourhood by the light 
 from bright field stars.  These   regions are indicated by white lines in  the right image, 
 where the occulting was not applied. 
 Red circles in the right  image mark  USNO stars used for astrometric referencing. 
 The yellow box region with the pulsar is enlarged in Fig.~\ref{fig:2}, \ref{fig:3}, and \ref{fig:4}.}
 \label{fig:1}
 \end{figure*}
\section{Observations and data reduction}  
\label{sec2}
\subsection{Observations}
The images of the pulsar field were obtained in the $V_{HIGH}$, $R_{SPECIAL}$, and $I_{BESSEL}$ bands 
with the FOcal Reducer and low dispersion Spectrograph (FORS2\footnote{For the instrument details 
see http://www.eso.org/instruments/fors/ }) at  the VLT/UT1 (ANTU) unit  during several service mode  runs 
in  April  2009. A high resolution mode was used with an image scale of  
$\sim$ 0\farcs126/$\mathrm{pixel}$  and a field of view (FOV) of $\sim$4\farcm2$\times$4\farcm2. 
Since PSR J1357$-$6429 is located near the Galactic plane 
($l$ = 309\fdg92, $b$ = $-$2\fdg51),
which is densely packed with stars, we used
the FORS2 MOS occulting-bar set-up for our observations.  This allowed us   
to minimise  contamination    
across the pulsar region  by both the illumination and saturation spikes of bright field stars 
surrounding the pulsar (see Fig.~\ref{fig:1}).     
Sets of three-to-ten minute non-dithered exposures were obtained in each of the bands.  
The observing conditions were rather stable and photometric, with  seeing values 
varying from 0\farcs4 to 0\farcs8.
The {\sl Log} of the observations is given in Table~\ref{t:log}.   
One short, ten-second exposure  was additionally taken in each of the bands without occulting bars 
(Fig.~\ref{fig:1}, right panel).  This was done to perform 
accurate astrometric  referencing:        
the exposure shortness minimises the saturation of most astrometric standards in the image 
and allows us to define their image positions with higher accuracy.   

We note that only 8 of 39 available individual 
exposures  of the best quality $I$-band data were used by  \citet{Mignani11}. 
This decreases the detection limit by at 
least a factor of two, which is crucial when searching for a faint  pulsar counterpart, and  
likely to be the main reason 
why  they missed the counterpart candidate  
clearly detected in the complete data set.      

Standard data reduction, including  bias subtraction, flat-fielding, and cosmic-ray
removal   was performed using the {\tt IRAF} and {\tt MIDAS}  tools.  
Preliminary data inspection showed  systematic frame shifts 
by a significant fraction of the pixel (of up to 0.1--0.2 pixels) between  
the first and  consecutive 
non-dithered exposures in each of the observing blocks typically 
consisting of 5--13 individual exposures.     
To correct for these systematic shifts,   
we  aligned all individual frames in  each of the bands, using a set of unsaturated stars, 
to a "best" quality frame for a given band 
obtained in the highest quality seeing conditions. The alignment accuracy was $\sim$ 0.01 of a pixel. 
In addition, we excluded four of the "worst"  exposures  in the $V$ band with seeing 
$\ga$ 0\farcs7, and two exposures  in  $R$ with seeing $\ga$ 0\farcs6.  
The resulting full width at half maximum (FWHM) of a point source  
on the combined $V$, $R$, and $I$ images was  $\approx$ 0\farcs59,  0\farcs52, and 0\farcs44, 
respectively, which is about 10\%--20\% better than for  a simple combining of initial frames without alignments.   
The  respective integration times are  12339.5 s,  7629.7 s,    and  7799.7 s, with mean airmasses of  
$\approx$ 1.347, 1.372, and 1.320.  
The pulsar was exposed on chip 1 of the FORS2 CCD mosaic consisting of two chips. 
An example of the combined $I$-band image for this chip is presented in the left panel of Fig.~\ref{fig:1}.  
Chip 2 was operated without  occulting bars, and even short images are strongly contaminated by several  
very bright and strongly over-saturated  field stars, making the data from this chip  useless  
for our goals.  
\subsection{Astrometric referencing} 
For astrometric referencing,  the  short VLT frame in  the $I$ band  was used (right panel of Fig.~\ref{fig:1}),
which is of higher  quality than 
the short exposures in other bands.  
To obtain a precise astrometric  solution,  the positions of the astrometric standards selected from the USNO-B1 
astrometric catalogue\footnote{USNO-B1 is currently
incorporated into the Naval Observatory Merged Astrometric Data-set
(NOMAD) which combines astrometric and
photometric information of Hipparcos, Tycho-2, UCAC, Yellow-Blue6, USNO-B,
and the 2MASS, http://www.nofs.navy.mil/data/fchpix/}
were used as  a reference. 
Thousands of  USNO-B1 reference objects can be identified in our crowded FOV.    
The recent release of the Guide Star Catalogue 
(GSC-II v2.3.2)\footnote{see http://gsss.stsci.edu/Catalogs/GSC/GSC2/GSC2.htm }
provides almost the same number of  standards  but contains 
no information on proper motions and the declared astrometric errors (0\farcs3) are
higher than the nominal 0\farcs2 uncertainty in the USNO-B1.
A number of brighter stars from the UCAC2 catalogue are also present  
but most of them are saturated in our images. We discarded the reference stars with significant 
proper motions and catalogue positional uncertainties $\ga$ $0\farcs3$ along with 
those that are saturated in our images.  
Finally, to minimise  potential  uncertainties 
caused by overlapping stellar profiles in the crowded FOV, we selected only 28 isolated stars 
marked by red circles in  the { \sl right panel} of Fig.~\ref{fig:1}. Their pixel coordinates 
were derived making use of the {\tt IRAF} task {\it imcenter} with the accuracy of $\la$ 0.025 of the image pixel. 
The {\tt IRAF} tasks {\sl ccmap/cctran} were 
applied to the astrometric transformation of the images.  Formal {\sl rms} uncertainties 
in the  astrometric
fit for our image  are $\Delta$RA $\la$ 0\farcs198 and $\Delta$Dec $\la$ 0\farcs181,
which are compatible with the nominal catalogue  uncertainty.
A selection of another set of isolated reference objects does not change the result significantly. 
The combined $V$ and $R$ images were preliminary aligned to the $I$ image 
reference frame with the accuracy of  $\la$ 0\farcs01.   
Accounting for that, a conservative estimate of our 1$\sigma$ astrometric referencing uncertainty is 
$\la$ 0\farcs2 in both RA and Dec for all three optical bands.

To check the \textit{Chandra} observation pointing accuracy for further 
comparison of the optical and  X-ray images,  
we also performed  astrometric referencing of the  J1357$-$6429 \textit{Chandra} archival 
X-ray images using optical astrometric catalogues.  Unfortunately, this was not possible for the  HRC-S images 
obtained with rather short exposures, where we found no suitable reference objects, even when  
we merged  two separate 
observational blocks (OBs)   
in one image,  to achieve a deeper  detection level. 
In a deeper  ACIS-I frame,  obtained with about twice as long an exposure,   
we found 18 point-like objects detected at $\ga$ 3.5$\sigma$ significance, 
which we identified with relatively bright optical reference objects  from the UCAC2 catalogue. 
Their image positions were defined with the  accuracy  $\la$ 0.5 of the ACIS  pixel size 
(pixel scale is $\approx$ 0\farcs5). 
The resulting {\sl rms} uncertainties in the respective  astrometric fit  are $\Delta$RA $\approx$ 0\farcs40 and 
$\Delta$Dec $\approx$ 0\farcs35 with maximal residuals $\la$ 0\farcs8.      
The  shift between the initial and transformed X-ray images was about 0\farcs15, 
which is insignificant within  the fit uncertainty.   
Our results are  compatible with a nominal \textit{Chandra} astrometric referencing uncertainty,    
ensuring us an almost perfect pointing accuracy for the X-ray observations.   
Below we  accept  the nominal uncertainty of 0\farcs6 as a conservative value for comparison 
with the optical data.   
    \begin{figure*}[t]
 \setlength{\unitlength}{1mm}
\begin{center}\includegraphics[width=105mm, clip=]{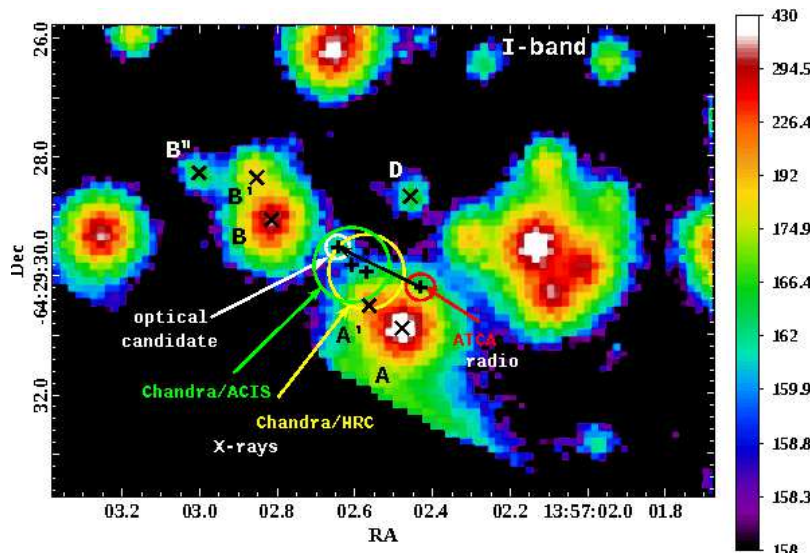}\end{center}
 \caption{Zoomed-in fragment 
 of the $I$-band image from the left panel of Fig.~\ref{fig:1}.  
 Black crosses and colour ellipses show the pulsar positions and their 1$\sigma$ 
 uncertainties in the early radio ATCA ({\sl red}) and  
 two later X-ray \textit{Chandra}  observations  ({\sl yellow and green}).  The optical counterpart candidate, 
 whose position is compatible with those in X-rays, is also indicated by the cross and   the white  
 uncertainty ellipse.  
 Black arrow shows possible pulsar proper-motion path 
 for  about a nine-year time base from the radio to the optical.  
 X-points and capital letters label  background sources located near/along this path.  
 A logarithmic scale is used 
 to enhance  the brighter and fainter objects. 
 The colour-bar identifies 
 the image values in 1000 counts units.    
    }
 \label{fig:2}
 \end{figure*}
\begin{table}[b]
\caption{Coordinates of PSR J1357$-$6429 measured at different epochs, in different ranges, 
and with different instruments.  
The optical data are for the optical counterpart candidate to the pulsar.
}
\begin{center}
\begin{tabular}{llll}
\hline\hline
   
    Date, range,       &  Epoch                                & RA$_{J2000}$     &  Dec$_{J2000}$                      \\
    instrument                          &  [MJD]              &[hms]   &  [dms]                    \\
   \hline \hline                
   29-08-2000, radio,        &             &                 &                                 \\
   ATCA$^a$               &  51785                   &   13 57 02.43(2)                         &  $-$64 29 30.20(10)        \\
     06-11-2005, X-ray,      &            &                                                     &                       \\
    \textit{Chandra}/HRC$^b$  &  53693                     &    13 57 02.57(9)                           &  $-$64 29 29.94(60)       \\                 
 25-04-2009, optical,             &            &                                                 &                          \\
    VLT/I-band$^b$           & 54946              &   13 57 02.64(3)                              &    $-$64 29 29.52(20)           \\ 
    09-10-2009, X-ray,              &            &                                              &                               \\
 \textit{Chandra}/ACIS$^b$           &  55113                  &   13 57 02.61(9)                                  &     $-$64 29 29.82(60)    \\ 
\hline
\end{tabular}
\begin{tabular}{l}
  $^a$ radio-interferometric position from \citet{Camilo04}    \\
  $^b$ this work; for the \textit{Chandra}/HRC epoch the coordinates are   \\
 compatible with those of \citet{Zavlin07}  \\  
\end{tabular}
\end{center}
\label{t:coords}
\end{table}
  \begin{figure*}[t]
  \setlength{\unitlength}{1mm}
 \resizebox{15.5cm}{!}{   
 \begin{picture}(150,130)(0,0) 
\put (45,88) {\includegraphics[width=95mm, clip=]{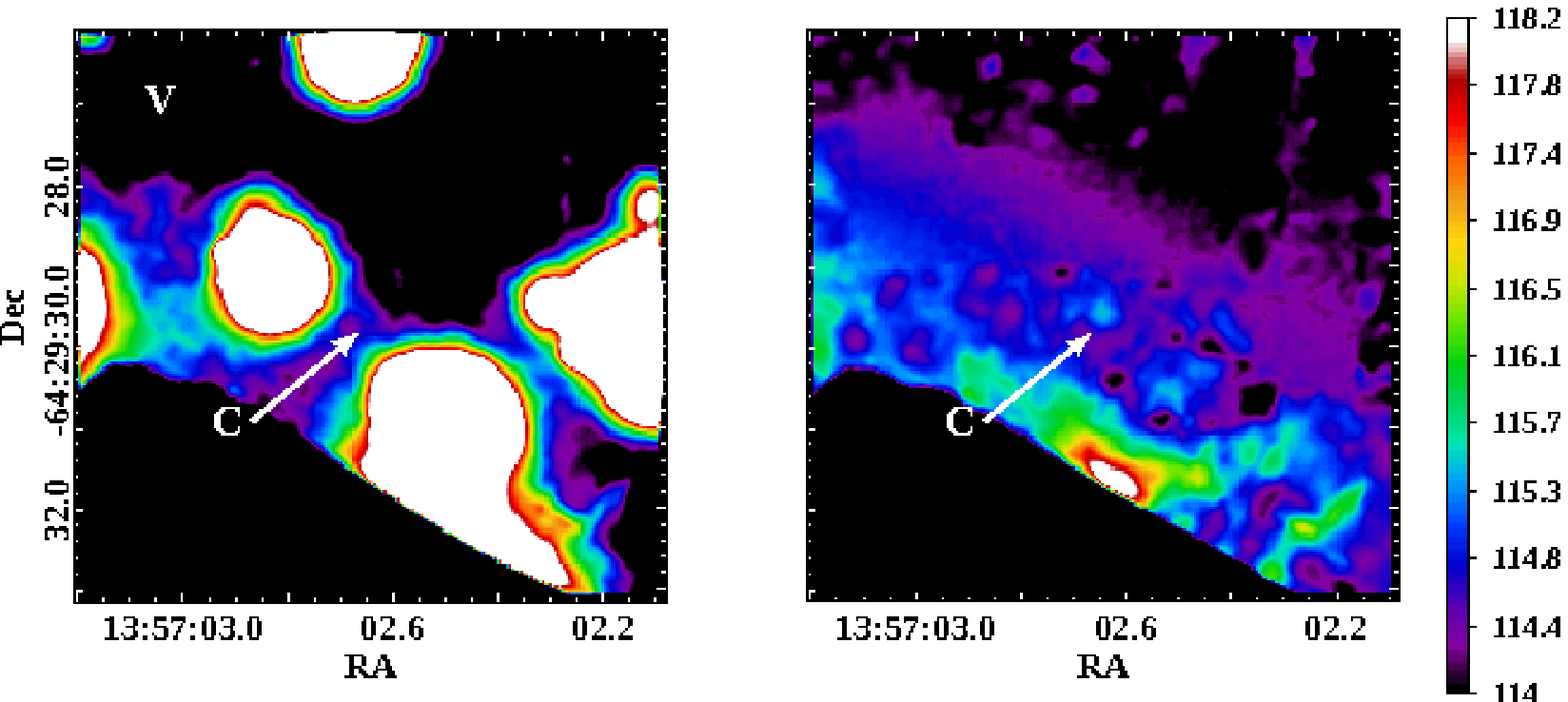}}  
 \put (45,44){\includegraphics[width=95mm,  clip=]{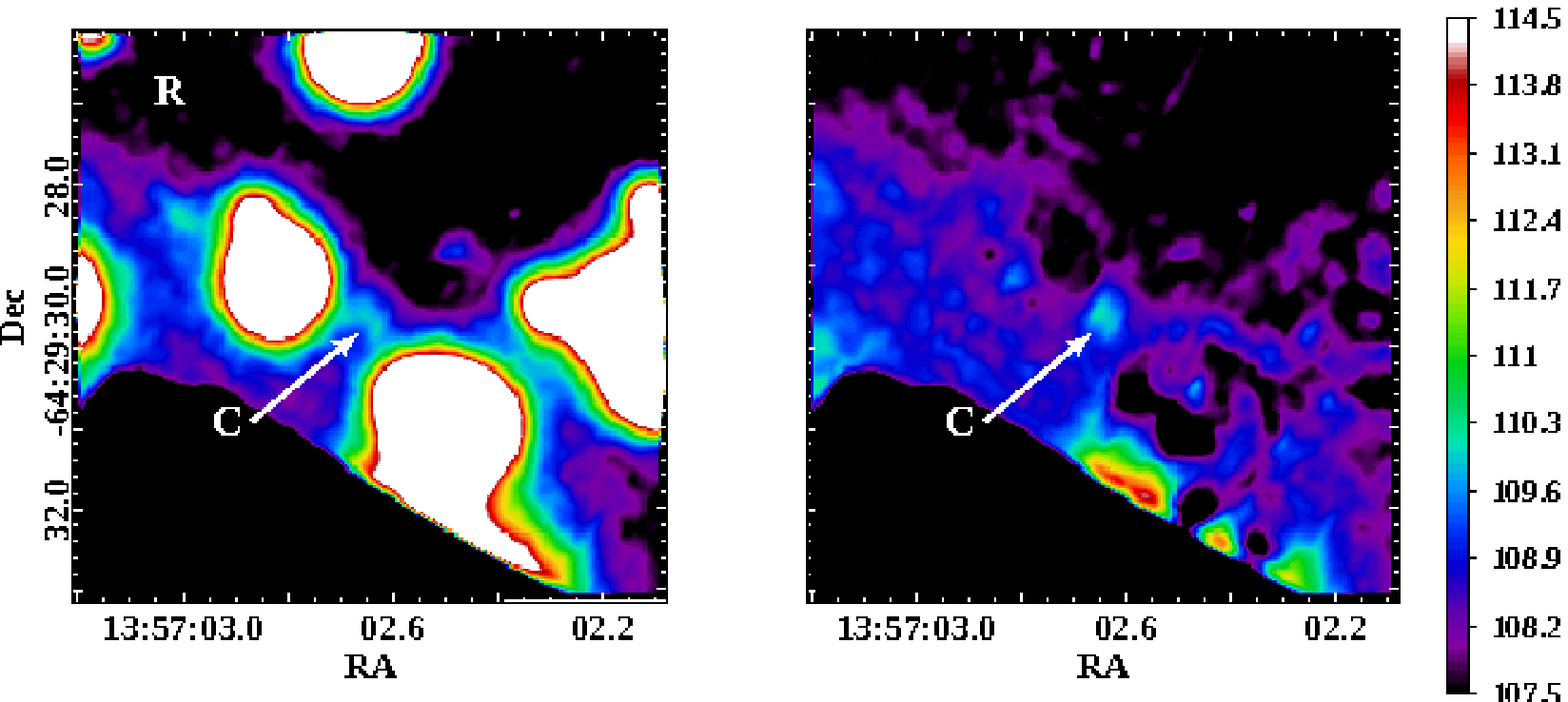}}  
 \put (45,00){\includegraphics[width=95mm,  clip=]{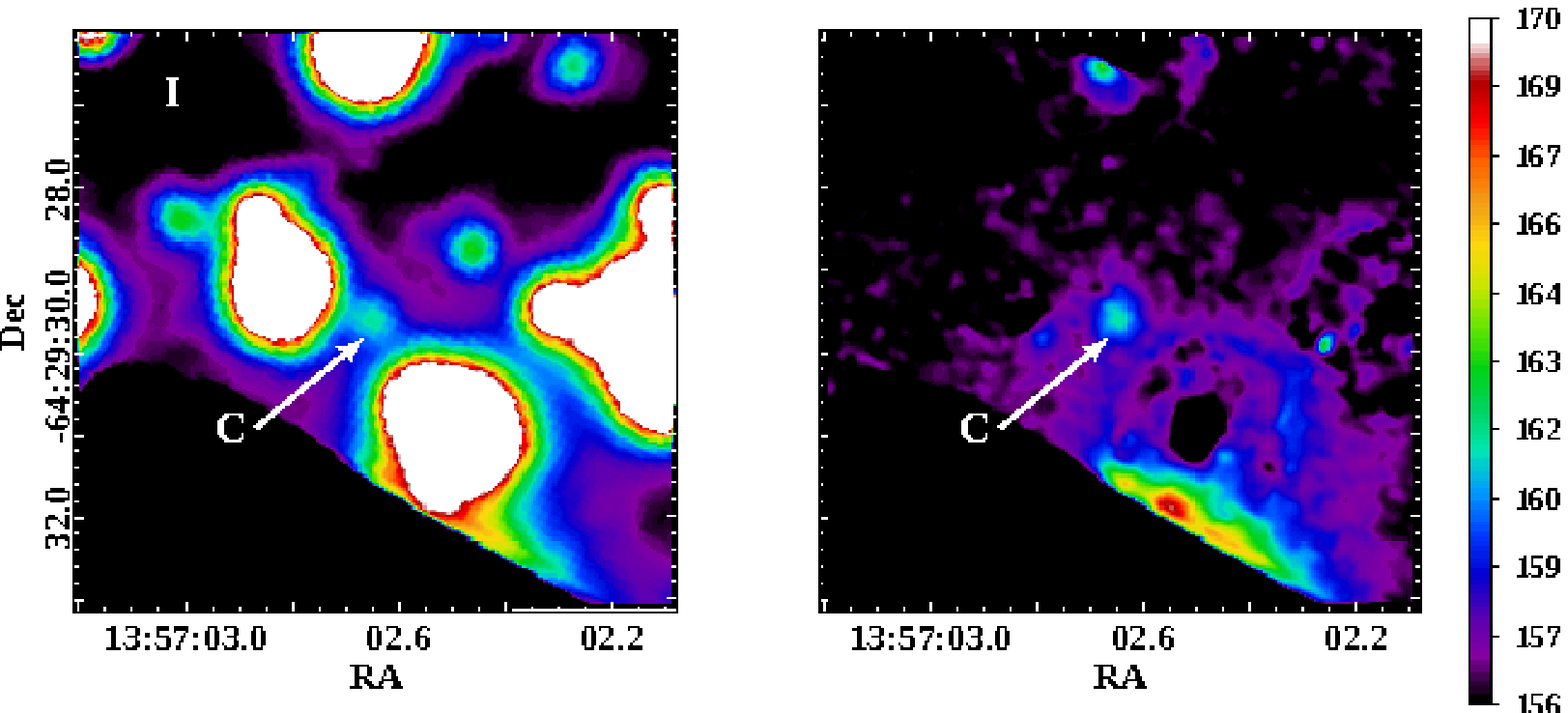}}    
  \end{picture}
      } 
 \caption{ 7\asec$\times$7\asec~fragments of the images of the PSR J1327$-$6429 field  
obtained  in   the $VRI$ optical bands  with the VLT. 
The band names are indicated in the top-left corners of the images. 
The right column panels show the respective images after subtraction of background stars, 
to better reveal the pulsar counterpart 
candidate labelled by C.  The sources overlapping with the south-east 
occulting boundary cannot be perfectly subtracted. 
The images are smoothed with 3 pixel Gaussian kernel reflecting mean  
seeing conditions. The candidate is firmly detected  in  $I$ and $R$ bands 
({\sl bottom and middle rows}) and only marginally 
resolved in $V$.  The colour-bar identifies the image values in 1000 counts units.     
   }
 \label{fig:3}
 \end{figure*}
\subsection{Photometric calibration}                
\label{photcal}
The observing conditions during our observations were photometric.
The photometric calibration was carried out with
standards from \object{E7}, \object{L107}, and \object{NGC2818} 
Stetson photometric standard
fields \citep{stets2000} observed during the same nights as
our target. We fixed the atmospheric extinction coefficients
at their mean values adopted from the VLT home page:
k$_V$ = 0\fm13 $\pm$ 0\fm01, k$_R$ = 0\fm075 $\pm$ 0\fm01, and k$_I$ = 0\fm056 $\pm$  0\fm01. 
As a result,  we obtained the following magnitude zero-points for
the summed images, $V^{ZP}$ = 28\fm14 $\pm$ 0\fm02, 
$R^{ZP}$ = 28\fm08 $\pm$ 0\fm03, and $I^{ZP}$ = 27\fm42 $\pm$ 0\fm03,  
and colour-term coefficients,  $VR_V$ =  0.03 $\pm$ 0.03,   $VR_R$ = $-$0.04 $\pm$ 0.04, 
and $RI_I$ = $-$0.01 $\pm$ 0.07,
where the errors account for the statistical uncertainties in the 
magnitude measurements, the extinction coefficient
uncertainties, and marginal  variations from night to night. 
The formal $3\sigma$ detection limits of a point-like object in the co-added images for 
a one-arcsecond aperture are $V$ $\approx$ 28\fm5, $R$~$\approx$~28\fm0, and $I$ $\approx$ 27\fm2.

\section{Results}
\label{sec3}
\subsection{Detection of the pulsar  counterpart candidate}
The $I$-band image fragment containing the pulsar  
is zoomed  in Fig.~\ref{fig:2}.  
The pulsar positions measured at different epochs 
in the radio and X-rays
are shown by crosses (+).  
Their 1$\sigma$ uncertainties are shown by ellipses,  
accounting for the 0\farcs2 uncertainty in the optical  astrometric referencing.        
The respective coordinates and observational epochs are 
listed  in Table~\ref{t:coords}, where  
the X-ray coordinates are remeasured by us.  
For the \textit{Chandra}/HRC observations, our results are consistent with those of \citet{Zavlin07}.  
As noted by \citet{Mignani11}, a significant difference between the radio  interferometric  
\citep{Camilo04} and X-ray pulsar 
positions  obtained at different epochs is indicative of a proper motion of the pulsar.   
From Fig.~\ref{fig:2}, one can see 
that  the  \textit{Chandra}/ACIS position is   
apparently shifted from  the \textit{Chandra}/HRC one,  
also in line with  the suggested motion.  
Although the shift is of a low significance, it  shows that the claimed proper motion 
may be real.       
  \begin{figure*}[t]
  \setlength{\unitlength}{1mm}
 \resizebox{15.cm}{!}{ 
 \begin{picture}(150,135)(0,0) 
\put (0,70) {\includegraphics[width=90mm, clip=]{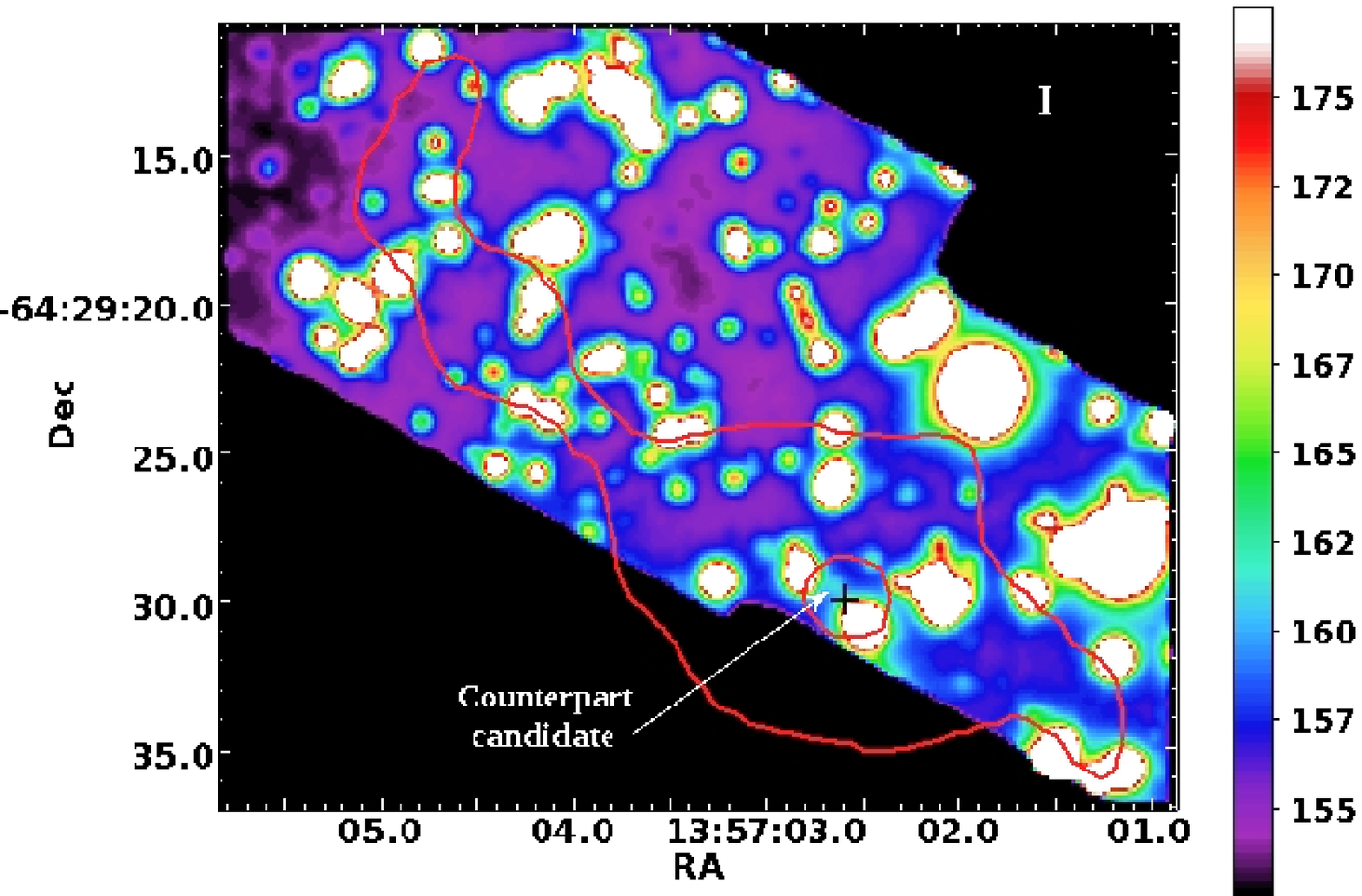}}  
 \put (94,70){\includegraphics[width=90mm,  clip=]{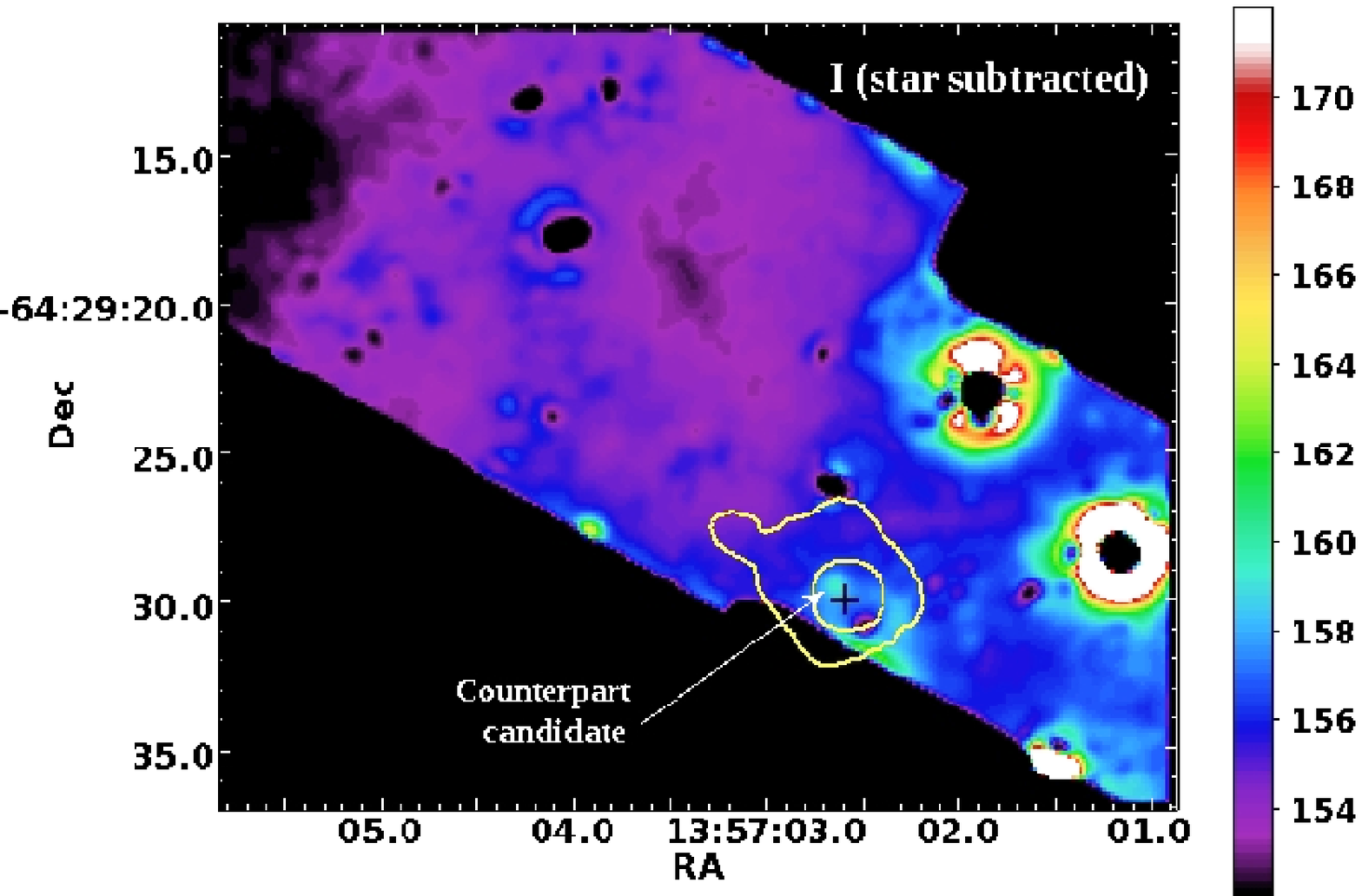}}  
 \put (0,0){\includegraphics[width=90mm,  clip=]{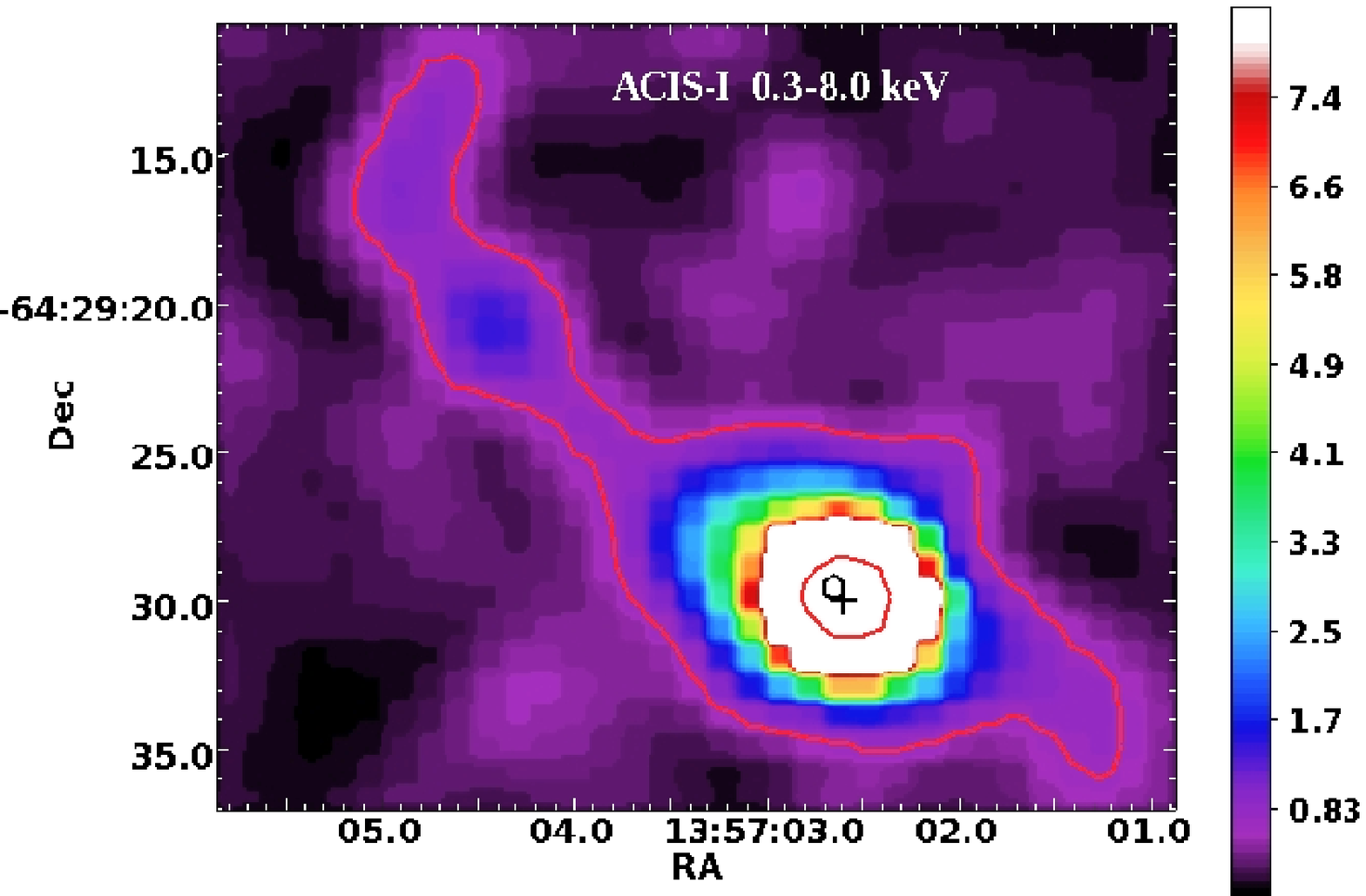}}  
  \put (94,0){\includegraphics[width=90mm,  clip=]{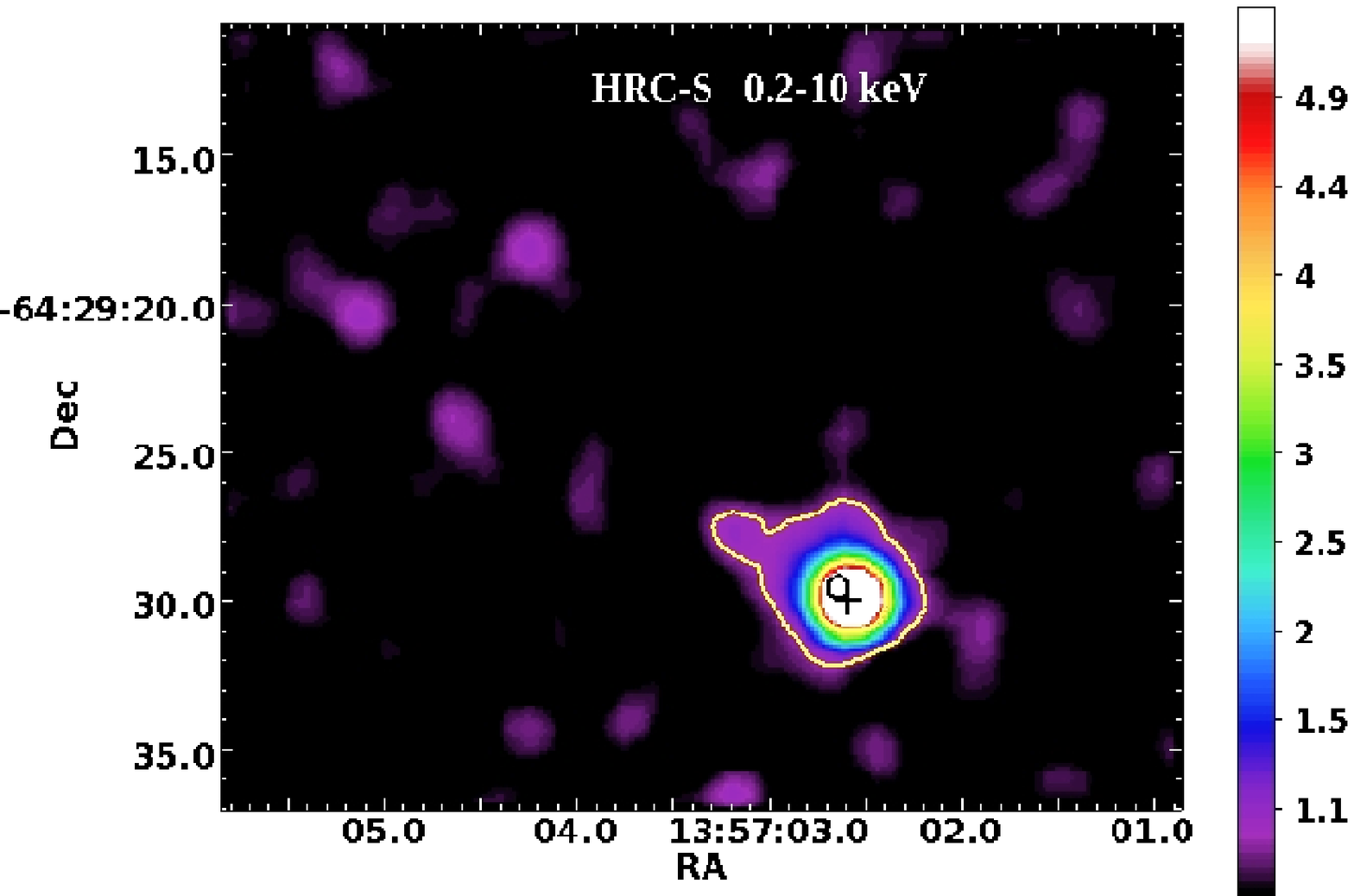}} 
  \end{picture}
  }
\caption{ Comparison of the fragment   of the optical $I$-band  
image of the PSR J1327$-$6429 field ({\sl top panels}) 
with   the corresponding 
fragments of the X-ray images obtained with \textit{Chandra} ACIS-I and HRC-S ({\sl bottom panels}).
The fragments sizes, $\sim$47\asec$\times$53\asec, are selected to show a tail-like 
structure of the   PWN extended NE of the pulsar and best resolved in 
the  ACIS-I image ({\sl bottom-left panel}). 
Only the brightest inner part of this structure is resolved in the 
HRC-S image   of a shorter exposure  
({\sl bottom-right panel}).  
The ACIS-I image is corrected by the exposure map, 
binned by two pixels, and smoothed with the three pixel 
Gaussian kernel. The  HRC image is binned 
by four pixels and smoothed with the same kernel. 
Background stars are subtracted 
in   the right optical panel. 
The optical images are smoothed with 
the three pixel kernel as well, that reflects   
the seeing value. The cross and its arms mark the X-ray position of the pulsar
and its 1$\sigma$ uncertainties obtained from the HRC image.     
The contours from the ACIS ({\sl red}) and HRC ({\sl yellow})  X-ray  
images of the pulsar/PWN system are overlaid on the optical left and right 
images, respectively. Internal contours show  
the FWHM of the ACIS and HRC PSFs, while the outer ones indicate 
the PWN  boundaries where this nebula merges with the background. 
The small black circle in the X-ray images shows 
the position and FWHM of the PSF of the faint pulsar optical counterpart 
candidate marked by an arrow. The candidate position is compatible with 
the X-ray pulsar position at the 1$\sigma$ significance level, 
while no extended PWN counterpart is visible in the optical. 
Colour bars show  the image values,  
which are in 1000 counts units for the optical and in single counts for X-rays.   
       }
 \label{fig:4}
 \end{figure*}

As seen from  Fig.~\ref{fig:2}, at least three of the 
optical objects  in the immediate vicinity of the pulsar 
can be considered as potential optical counterparts of the pulsar. 
These are  the objects $A$, and  $A'$ marked by   X-points, 
and the {\sl "optical candidate"}  marked by the plus. 
The brightest one,  object  $A$, was analysed 
and rejected as the counterpart by  \citet{Mignani11}. 
Its colours suggest that this is  a   main sequence star.    
The fainter object $A'$, which overlaps with $A$ and lies closer to the X-ray position of the pulsar,
has not yet been analysed.  Only a    {\sl "background enhancement"}, which has been difficult 
to identify with a real object, was detected at the position 
of the  "optical candidate" 
in the five times shorter exposure image analysed   by  \citet{Mignani11}. 
In our  image, this source is clearly detected 
at  $\ga$ 10$\sigma$ significance level and is certainly 
not the result of the overlapping wings of the nearby brighter stars 
$A'$ and $B$.   Among the  other potential candidates, 
its position (presented in Table~\ref{t:coords}) 
is  most closely  compatible with the X-ray position of the pulsar obtained  from  
the ACIS-I observations taken in the same year as the optical data. 
It is also compatible  with the path of the potential 
proper motion of the pulsar marked in  Fig.~\ref{fig:2} by a black arrow. 
This allows us to  consider this source as 
the most plausible  candidate for  the optical counterpart of the pulsar. 
 
If  this is a real counterpart, the respective proper motion of the pulsar, 
based on about  8.7 yr time base between the radio and optical observations, 
is    $\mu$ = 177 $\pm$ 37 mas/yr, with a position  
angle 67\grad $\pm$ 8\grad. This is compatible with  estimates based 
on the difference between the radio and X-ray positions 
of the pulsar \citep{Mignani11}.  The significance of our result, 
$\sim$ 4.8$\sigma$,  is  higher  
than the 3$\sigma$  quoted by \citet{Mignani11} 
because  the optical source is localised to higher accuracy. However, the real nature   
of the source is still to be  examined.   
In addition to   the $A$ and $B$ stars, which were analysed and rejected as counterparts  
by  \citet{Mignani11},  below we analyse  in detail the magnitudes and colours  
of the optical candidate,  
the object $A'$,  and   the   point sources, 
which are  marked by X-points and located close to the pulsar position 
and/or its possible proper motion path.  
    
In Fig.~\ref{fig:3}, we compare the $VRI$ images of the same field   
and demonstrate that
the optical candidate, which   is marked  by a $C$ there, 
is also detected in the  $R$ and $V$ bands. However, it becomes fainter   
and is detected at $\sim$ 7$\sigma$ and  $\sim$ 3$\sigma$ levels, respectively,  
which is considerably lower than in the $I$ band. 
To help identify this source, we constructed  a model point spread function (PSF) for each  band   
based  on a set of about 30 isolated and unsaturated field  stars  and used 
the {\sl allstar} task from the {\tt IRAF} {\sl digiphot} tool. 
The task provides an efficient iterative algorithm for a careful subtraction  of  groups of stars with 
overlapping PSFs. The latter  is crucial in our case, since most of the stars near the pulsar   
are in  these groups.  The  subtraction process is  somewhat imperfect,  and  
we  repeated it  several times varying the {\sl allstar} parameters. The results for each of the bands 
were rather stable, and  they are shown in  the {\sl  left panels} of Fig.~\ref{fig:3}.
The main contaminating factors for $C$,  
the nearby objects $A$, $A'$, and $B$ (cf.~Fig.~\ref{fig:2}),  
are  perfectly subtracted. 
This is obviously  impossible for  saturated stars  and/or stars 
whose images overlap 
with the occulting boundary.  
Although their  residuals
are noticeable,  they are  distant enough from the optical candidate  
and cannot   affect  its flux considerably.  
  
The star-subtracted images confirm that in all three bands we indeed detect 
the same source $C$  that we have suggested is the most 
plausible candidate for the pulsar optical counterpart.  
We have carefully analysed  individual exposures in all three bands 
and found that this source   is  not an artefact 
of either a poor cosmic-ray  and  bad CCD-pixel rejection   or inadequate flat-fielding. 
The significance of its detection 
is   higher in the star-subtracted images with smoother backgrounds.  
From the comparison of the $VRI$ images, we also see 
that $D$ is a very red object, which is not detected in the $V$ band, 
and $B''$ is also hardly resolved in this band. 

\subsection{Searching for a PWN  counterpart}
In Fig.~\ref{fig:4}, we compare the optical and X-ray images 
of the pulsar+PWN system on a wider spatial scale  
including the X-ray PWN tail-like extended structure in the immediate vicinity 
of the pulsar. It is most clearly visible with the \textit{Chandra}/ACIS ({\sl bottom-left panel}). 
Only the brightest part of the structure 
is seen in a shorter exposure image obtained earlier with the \textit{Chandra}/HRC ({\sl bottom-right panel}). 
As seen, the  pulsar counterpart candidate 
position in the $I$ band is fully compatible with the X-ray position of the pulsar, 
while  no  reliable counterpart to the extended X-ray structure is visible  
even in the star-subtracted optical image ({\sl top-right panel}).   
The same is true for the $V$ and $R$ images that are not shown here.   

The similar situation is known for the Vela PWN, 
which is not detected in the optical  in spite of  numerous attempts \citep{mig03,shib03,Danila11}, 
while the Vela-pulsar optical counterpart is a rather bright and firmly detected point-like  object.   
\subsection{Photometry}
To analyse the  magnitudes and   colours of the potential optical counterparts  and sources marked  
in Fig.~\ref{fig:2},   
and to compare  them with
field stars, we performed  the PSF photometry   
on the summed images  using 
the outputs of the  {\sl allstar}  task, which we used for the star subtraction. 
The PSFs were generated for nine-pixel radius, where the bright isolated unsaturated stars 
selected for the PSF construction merge with the background.    
An optimal PSF fit radius  for field sources was chosen  to be about three, four, and five pixels  
for the  $I$, $R$, and $V$ bands, respectively,  
in accordance with different 
seeing values in these bands (Sect.~2.1).     
Annulus and dannulus  of nine and ten pixels were typically used   to extract local  backgrounds. 
They were  smaller for $A$, $A'$,  $B$, $B'$, $B''$, and the optical candidate $C$,  to escape  
a false background determination near the occulting boundary.   
The respective aperture corrections were estimated and applied based 
on the photometry of bright unsaturated field stars.  
We also performed the aperture photometry. 
Although  the results were consistent with the PSF ones, 
the magnitude errors were typically larger than those for the PSF photometry, particularly for fainter objects. 
Therefore, below we present 
more  accurate  PSF photometry results, which are collected in Table~\ref{t:mag}.  
The errors include  the statistical  measurement errors, 
calibration zero-points uncertainties,   and  corrections for  possible star-subtraction uncertainties 
based on a magnitude dispersion after  the reiterations of the  subtraction process. 
Our colour-term calibration coefficients are sufficiently small (see Sect.~\ref{photcal}), and any 
colour-term contributions  
in the transformation from instrumental  to real magnitudes for each of the objects and bands   were evaluated  
to be  insignificant. They were always within  the estimated  error budgets  even for the reddest  
field object $D$  with $V-R \ga 2$ and $R-I \approx 2.5$. We ignore them in further considerations.           
The stellar magnitudes were transformed into fluxes  using zero-points provided by \citet{Fukugita}. 
\begin{table}[t]
\caption{ Measured $VRI$ magnitudes and fluxes $F_{Band}$ (in $\mu$Jy)  of the point-like optical 
sources detected in the PSR J1357$-$6429 vicinity and labelled in Fig.~\ref{fig:2}. Numbers in brackets are
1$\sigma$ uncertainties referring to the last significant digits quoted.}
\begin{center}
\begin{tabular}{llll}
\hline\hline
 Source               &  $V$                &  $R$       &  $I$                    \\
                  &     $F_V$            &  $F_R$                &   $F_I$                            \\
   \hline \hline                
   $A$     &    21.70(02)           & 20.72(03)           &   19.77(02)                                  \\
               &    7.5(2)             &  15.5(4)          & 29.5(5)    \\ 
                 &                 &            &     \\ 
     $A'$       & 24.48(03)             &   23.22(04)          &    22.02(03)                    \\
              &     0.58(02)             &   1.56(05)          &  3.7(1)   \\ 
	                 &                 &            &     \\      
   $B$     &  22.79(02)           &   21.67(03)         &     20.45(02)                \\                             
     &     2.75(05)              &    6.5(2)          &  15.8(3)   \\ 
                   &                 &            &     \\ 
   $B'$     &  25.58(05)        &  24.18(04)          &          22.56(02)                 \\
     &   0.210(009)               &   0.64(02)           &  2.26(05)  \\ 
                     &                 &            &     \\ 
    $B''$      &  27.4(4)          & 25.51(08)          &    23.74(03)                  \\ 
      &  0.04(01)               &    0.19(01)         &  0.76(02)   \\ 
                    &                 &            &     \\ 
    $C$ (counterpart     &    27.3(3)          & 25.52(07)             &   24.13(05)  \\ 
     candidate)  &     0.04(01)             &    0.19(01)            &  0.53(02)   \\ 
                    &                 &            &     \\ 
    $D$      &   $\ga$ 28.1                  &   26.1(1)          &   23.67(04)  \\ 
      &  $\la$ 0.02               &     0.106(009)        &  0.81(03)   \\ 
\hline
\end{tabular}
\end{center}
\label{t:mag}
\end{table}
 \begin{figure*}[t]
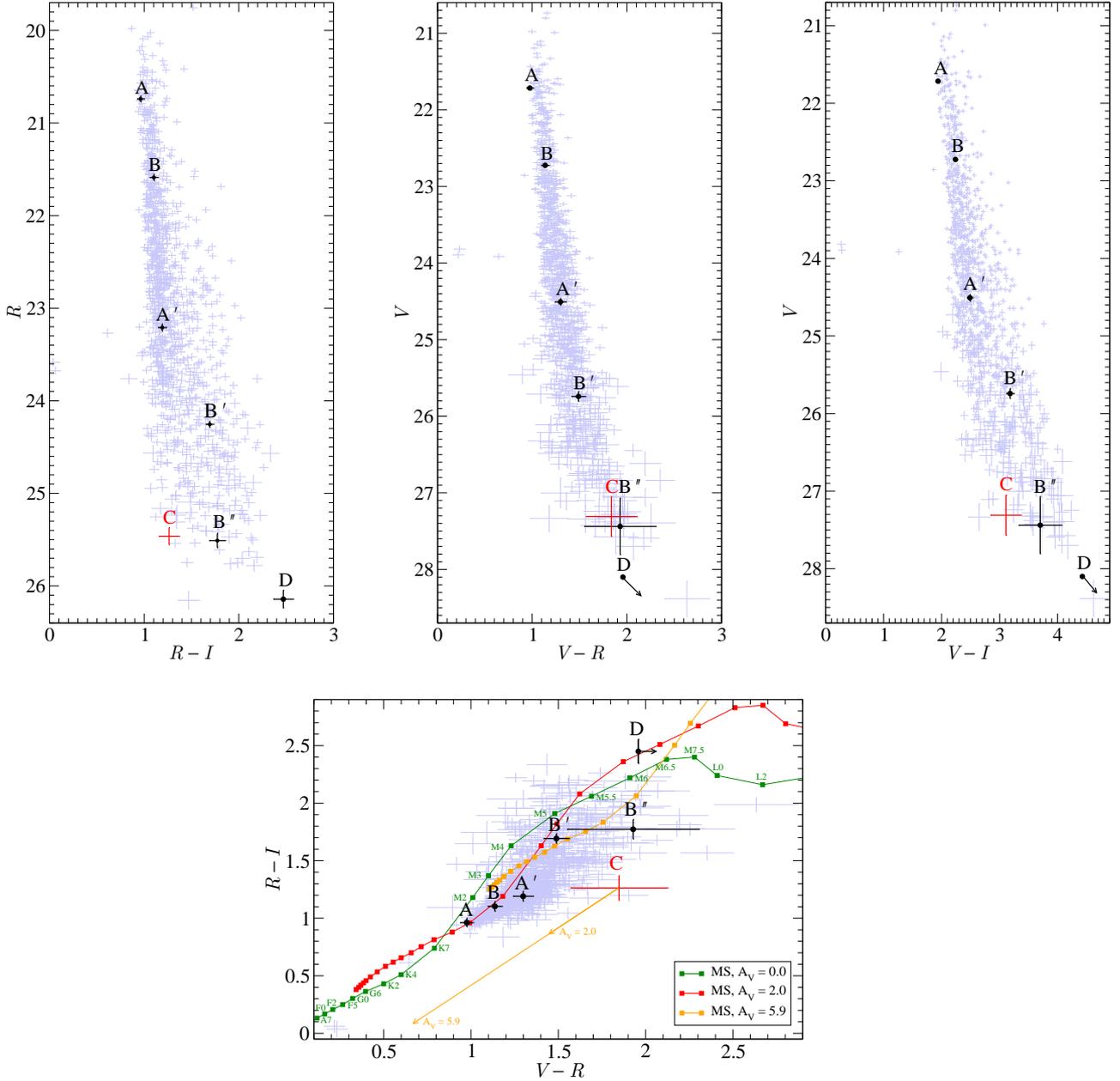

  \setlength{\unitlength}{1mm}
 \resizebox{18.cm}{!}{ 
 \begin{picture}(180,170)(0,0) 
\put (5,65) {\includegraphics[scale=0.35, clip]{R_vs_R-I.eps}}   
 \put (65,65){\includegraphics[scale=0.35, clip]{V_vs_V-R.eps}}   
 \put (125,65){\includegraphics[scale=0.35, clip]{V_vs_V-I.eps}}   
  \put (45,0){\includegraphics[scale=0.35,clip]{V-R_vs_R-I.eps}}  
  \end{picture}
  }
 \caption{ The observed colour-magnitude ({\sl top panels}) and colour-colour ({\sl bottom panel}) 
diagrams of PSR J1357$-$6429 field stars.
The counterpart candidate labelled by C is shown by red, and nearby stars marked in Fig.~\ref{fig:2} 
are shown by black.  Empirical  main-sequence colour-colour curves reddened with  a zero,  medium, and  
entire Galactic extinction, $A_V$ = 0, 2.0, and 5.9,
are shown at the {\sl  bottom panel} by green, red and yellow lines, respectively.  
The spectral classes are  marked by filled squares and indicated  along the green curve.   
The extinction vectors in the {\sl bottom panel} show where  
C-object should move after dereddening with  $A_V$ = 2.0  or  5.9.      
       }
 \label{fig:5}
 \end{figure*}
\subsection{Colour-magnitude and colour-colour diagrams} 
Fig.~\ref{fig:5} presents the observed colour-magnitude and  colour-colour diagrams, collecting  
about one thousand  point-like field objects   detected  in  a magnitude range  of $\sim$ 19--28. 
As in the previous section, the PSF photometry based on the outputs of the  {\sl allstar}  task 
was used to estimate the magnitudes.   
To minimise the appearance of  blended doubles and partially resolved galaxies, 
only the objects with the {\sl allstar} sharp parameter $\la$~1 detected at  
$\ga$~4$\sigma$ point source significance  were selected.    
Brighter objects   are not included, because  they are saturated  in long integration-time images 
and  cannot be measured reliably.  This is not critical to our analysis, 
since all the objects of interest,  which are  marked in Fig.~\ref{fig:2}  
and listed in Table~\ref{t:mag}, 
are in the considered  magnitude range. 
All of them are      labelled  in Fig.~\ref{fig:5}. 

Some of the diagrams for a  magnitude range of 15--25 were presented by   
\citet{Mignani11}.  To avoid the saturation problem for brighter objects,
they used   only short ten-second exposures,  
but  these exposures had about a three-magnitude lower detection limit. As a result,  
their colour-magnitude diagrams at magnitudes $\ga$~22 have a significantly larger 
colour dispersion than our ones, and, therefore, 
they are less informative for the colour analysis of fainter sources, including 
most  of the objects marked in Fig.~\ref{fig:2}.        
  
The main stellar sequence distribution  is clearly resolved  up to  $\sim$ 28
visual   magnitude level in our  colour-magnitude diagrams, 
which have a sharp edge   
at a low colour-index limit. This means that we have collected a considerable 
portion of the  Galactic stellar objects 
of the considered magnitude range  located within an arcminute 
width beam shaped  by the occulting mask and   
directed toward the pulsar.  
Only  a few objects 
with $V$ $\approx$ 24 
are shifted   off  the distribution  
towards a zero colour position.  They are located far away from the pulsar 
position and  may be  either  white dwarfs or  extragalactic  objects.  

In the colour-colour diagram, we also show empirical main-sequence curves \citep {bessel1979,bessel1991} 
extended   towards brown dwarfs    \citep{dahn2002} and  reddened with  a zero, medium,  and 
entire Galactic extinction, $A_V$ = 0, 2.0, and 5.9 \citep{schleg98}. 
The colour-colour positions of all the objects marked in Fig.~\ref{fig:2}, 
except C, can be linked  with the empirical curves 
by an appropriate  choice of $A_V$.  
For instance, the brightest objects $A$ and $B$  from the nearest 
vicinity of the pulsar lie almost entirely on the main sequence branch reddened with $A_V$ = 2.0.  
This allows us to roughly identify them 
as K4V and K6V spectral type stars,  respectively,  
at the  distance  of about 4.5 kpc. 
This confirms with a higher confidence  
the early conclusion  \citep{Mignani11} that   they are ordinary 
background stars. 
  
The   object $A'$, that is  fainter and located   closer to the pulsar, as well 
as the more distant $B'$,  are also consistent with the  main  sequence distribution. 
The colour properties of  a fainter $B''$ are less certain,  but it is likely compatible  
with the observed main-sequence behaviour at a high-magnitude/high-colour/high-extinction limit.   
The faintest object $D$ is  an extremely red object,  probably a brown dwarf. 
The colour information for all these sources  shows that they are background objects 
unrelated to either the pulsar or  its PWN.  

The colour properties of the  counterpart candidate $C$  
are likely   distinct  from those of field  stars. This is  most definitely 
seen in the $R$ versus (vs) $R-I$ diagram, where its position  uncertainties are significantly 
smaller than those in other colour-magnitude plots, 
and where it  is apparently shifted to the left from the main sequence 
distribution,  which bends to the right at  $R$ $\ga$ 24.    
The  colour differences  between the candidate and field sources 
in the $V$ vs $V-I$ and $R-I$ vs $V-R$ diagrams  are also visible, 
though they are less certain owing to the large uncertainties in the $V$ magnitude 
of the candidate. 

The position of this source in all the diagrams  does not allow 
us to definitely 
discard it  as a counterpart, as was done 
for the $ABD$ objects.   Any reasonable dereddening,  
shown by the extinction vector  
in   the $R-I$ vs $V-R$ diagram,  cannot place it in the main sequence branch.  
Therefore, its  unusual colour properties 
are atypical of  
a field star but  compatible with some other pulsar optical 
counterparts usually detected as  faint blue objects.  
For instance,  a similar position relative to the distribution 
of field stars in the $R$ vs $R-I$ and $V$ vs $V-R$ diagrams  
is occupied by  the Geminga-pulsar optical counterpart   
\citep{Kurt01,shib06}.  This and  the  positional  agreement 
of the source $C$ with the pulsar, allows us to 
keep it as the only likely  optical counterpart 
of the pulsar J1357$-$6429. 
\begin{table*}[tbh]
\caption{  Absorbed  BB$+$PL and NSA$+$PL fits to the  the pulsar and pulsar$+$PWN spectra observed  
with the \textit{Chandra}/ACIS-I and \textit{XMM}. 
 }
\begin{center}
\begin{tabular}{llllllll}
\hline\hline
\multicolumn{8}{c}{ }  \\ 
 $N_H$              &        $T$       &  $R^2_{km}/$ 
 & $\Gamma^{PSR}$   & PL$_{norm}^{PSR}$    &  $\Gamma^{PSR+PWN}$    & PL$_{norm}^{PSR+PWN}$  & $\chi^2$/dof (dof)              \\ 
                 10$^{21}$  cm$^{-2}$  &  keV          &   $D^2_{10kpc}$         &     &  10$^{-6}$ph cm$^{-2}$ s$^{-}$ keV$^{-1}$  &     &  10$^{-6}$ph cm$^{-2}$ s$^{-}$ keV$^{-1}$  &    \\
   \hline \hline
\multicolumn{8}{c}{ }  \\ 
                  \multicolumn{8}{c}{BB$+$PL}  \\ 
                       \hline  
		        &                 &            &    & & & &  \\ 
   1.52$^{+0.66}_{-0.56}$     &   0.190$^{+0.026}_{-0.026}$            & 3.8$^{+3.4}_{-3.4}$   &    1.50$^{+0.27}_{-0.34}$  &   8.89$^{+4.57}_{-3.99}$  &  1.26$^{+0.12}_{-0.12}$ & 17.0$^{+2.8}_{-2.8}$ &  0.9599 (167)                             \\
                        &                 &            &    & & & &  \\ 
   3.5      &   0.140$^{+0.007}_{-0.007}$            & 50.9$^{+14.4}_{-14.4}$             &    1.73$^{+0.19}_{-0.13}$  &   13.0$^{+4.9}_{-4.2}$  &  1.46$^{+0.08}_{-0.08}$ & 23.0$^{+2.3}_{-2.3}$ & 0.9981  (168)                             \\
   frozen                    &                 &            &    & & & &  \\ 
\hline 
\multicolumn{8}{c}{ }  \\ 
                    \multicolumn{8}{c}{NSA$+$PL}      \\ 
\hline 
                        &                 &            &    & & & &  \\ 
  3.44$^{+0.24}_{-0.17}$   &   0.079$^{+0.001}_{-0.001}$            & 1600             &    1.60$^{+0.24}_{-0.13}$  &   10.8$^{+2.8}_{-3.0}$  &  1.39$^{+0.08}_{-0.08}$ & 20.8$^{+4.1}_{-4.1}$ &  0.9669 (168)                             \\  
                         &                 &     frozen       &    & & & &  \\   
\hline
\end{tabular}
\end{center}
\label{t:x-fit}
\end{table*}
\subsection{Multi-wavelength spectrum}
Using the measured optical fluxes of the pulsar  counterpart candidate together  
with the available X-ray data, it is useful to construct a tentative multi-wavelength 
spectrum of the pulsar and compare it with the spectra of other pulsars detected 
in  both spectral domains. 
 \subsubsection{X-ray spectral analysis} 
As  shown by \citet{Chang2011} and  \citet{Lemoine-Goumard11} 
based on \textit{XMM-Newton} and \textit{Chandra}/ACIS 
observations, the X-ray spectrum of the pulsar can be described by  either 
a  blackbody-plus-power-law  (BB$+$PL) 
model or the  magnetized neutron-star atmosphere plus power-law (NSA$+$PL) model.   
Owing to the
lower spatial resolution,  the flux of the point-like pulsar  
in the \textit{XMM} data is contaminated significantly  by the extended 
emission from the PWN, which also has a PL spectrum, 
while  the pulsar is clearly resolved
in the \textit{Chandra}/ACIS data. 
For any model,  the thermal spectral component  (BB or NSA), presumably 
originating from the surface of the  neutron star (NS),  
strongly dominates the   emission detected with both instruments  
in  soft and middle energy channels
($E$ $\lesssim$ 1.5 keV). 
The non-thermal PL component from the pulsar  contributes significantly 
only to the high energy channels, 
but it is  impossible to distinguish it from the PL component of the PWN 
in the \textit{XMM} data. 
   
Accounting for  that, we performed independent X-ray spectral fits.  
We separated the data into two groups:   
the \textit{XMM}/EPIC/MOS spectra  of the pulsar+PWN system extracted 
from 20\arcsec circular aperture  compatible with the 
\textit{XMM} PSF and centred on the pulsar;  and the \textit{Chandra}/ACIS-I  spectrum 
of the pulsar extracted from a 1\farcs5 circle aperture compatible with the ACIS PSF. 
To determine the model parameters more accurately using the data from both instruments, 
we fitted both groups simultaneously  using the {\tt Xspec  v.12.3}.  
In our fits,  the absorbing column density $N_H$ and  the thermal component of the model (BB or NSA)  
were defined as common   for both groups, while the power law components were  defined individually 
for each of the groups  to determine the  parameters for the pulsar and 
pulsar$+$PWN nonthermal components, separately.  
As in  the  papers  cited above, 
for the NSA model we  fixed the NS mass at 1.4\msun, radius at 10 km, 
magnetic field  at 1 $\times$ 10$^{13}$ G, estimated from timing observations, 
and the distance at 2.4 kpc based on  
the DM  \citep{Camilo04}. The data were fitted 
in the 0.2--10 keV range, and the results are presented in Table~\ref{t:x-fit}. 
Our results are consistent with the published ones
within the uncertainties.  
At the same time, the parameter  
$N_H$ and the pulsar photon index   $\Gamma^{PSR}$ for each of the models  
appear to be less uncertain 
than the 
published fits. This is important for further 
comparison to the optical data.
We also note, that  the
best-fit  $N_H$ value is about twice as small 
for the BB$+$PL model  as
for NSA$+$PL,  although  "freezing" $N_H$  at  
the NSA$+$PL value provides only a marginally poorer BB$+$PL 
fit for which the NS surface temperature $T$  is 
considerably lower and the radius of the emitting 
area $R$  is larger than the corresponding values obtained at the "thawed" $N_H$.  
The NSA value of    
the column density may be more consistent 
with the DM value for the pulsar \citep{Zavlin07}. 
However,  as  similar significances of 
the  spectral fits do not allow us to conclude which of the models 
(NSA or BB) is closer to reality,  we 
compare our optical data with both models.  

To do that, one has to correct the observed  optical fluxes for the interstellar extinction.  
However, the extinction value  $A_V$ to the pulsar  is unknown. 
It can be roughly estimated from  an empirical
$A_V$--$N_H$ relation \citep{pred95},  using  $N_H$ values obtained  above from  the X-ray fit.   
It can also be additionally constrained  using the  DM    
distance to the pulsar and the so-called  red clump stars from 
the pulsar field as standard candles  
for obtaining the $A_V$--distance relation toward the pulsar. Both methods can be combined 
to provide  an independent  distance to the pulsar 
estimates  and additional constraints on $A_V$ and $N_H$. 
  \begin{figure}
\begin{center}
\includegraphics[scale=0.35, clip]{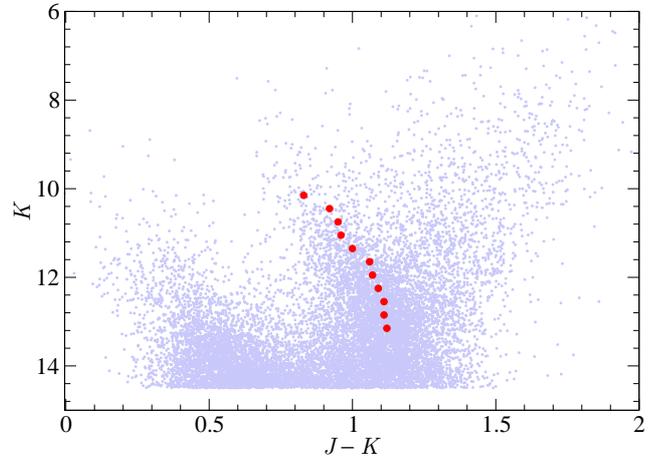}
\end{center}
 \caption{Near-infrared colour-magnitude diagram of the stars located within 0.3 deg of the pulsar position 
 ($l$ = 309\fdg92, $b$ = $-$2\fdg51), taken from the 2MASS All-Sky Point Source Catalogue.
 Filled circles point to the mean positions of the red clump stars for each magnitude bin (see text). 
 }
 \label{fig:2MASS-diag}
\end{figure}
 \subsubsection{Red clump stars and the $A_V$--distance relation} 
To derive the $A_V$--distance relation toward the pulsar,  we followed a method described 
in \citet{lopez2002}, \citet{cabrera2005}, and \citet{guver2010}.

Firstly, we extracted from the  2MASS All-Sky Point Source Catalogue
\footnote{see http://irsa.ipac.caltech.edu/applications/DataTag/, 
DataTag $=$ ADS/IRSA.Gator\#2011/0530/083402\_32642}  
about 17000  stars which are located within 0.3 degrees of the pulsar position 
($l$ = 309\fdg92,  $b$ = $-$2\fdg51) and created  the  colour-magnitude diagram,  
$K$ vs $J-K$ (Fig. \ref{fig:2MASS-diag}). The main sequence ({\sl left}) and red clump ({\sl right})  branches  
are clearly resolved in this diagram. We  divided the red clump branch   into several magnitude bins in the range 
of 10 $<$ $K$ $<$ 13.3 with a bin size of $d$K = 0.3.  In  each of the bins, we constructed 
a histogram of the star  
distribution    over  the $J - K$ colour  and found  a mean  colour by  fitting the histogram with  a Gaussian. 
The mean colours for all the bins  are  marked by  filled circles in Fig. \ref{fig:2MASS-diag}.  

\begin{figure}
\begin{center}
\includegraphics[scale=0.3,clip]{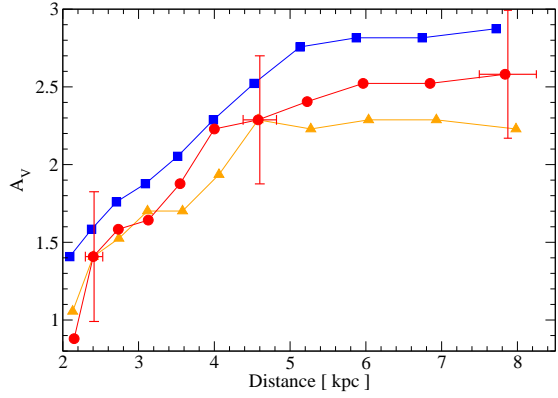}
\end{center}
 \caption{The  derived  $A_V$--distance relations    for the line of sight toward  the pulsar  
 (red) and for lines  shifted by $+$0.5 
 (blue)  and  $-$0.5 
(orange)  degrees  away from the pulsar along the Galactic longitude.  
}
 \label{fig:Av_vs_dist_comp}
\end{figure}

Secondly, for the red-clump population we  accepted  an absolute magnitude  
M$_{K}$ = $-$1.62 $\pm$ 0.03, 
and   intrinsic colour  ($J- K$)$_{0}$ = 0.68 $\pm$ 0.07 (\citet{lopez2002}, 
\citet{cabrera2005}, \citet{guver2010}). 
Using that and the above colour fits,  for  each of the bins we calculated  the K-band extinction,  
$A_{K}$ = $c_{e} \left[(J-K)-(J-K)_{0}\right]$, where $c_{e}$~= 0.657 \citep{rieke1985},  
and the distance  D = 10$^{(m_{K}-M_{K}+5-A_{K})/5}$ pc. 
Combining the results for all the bins, and using the relationship $A_{K}$ = 0.112$A_{V}$ \citep{rieke1985},  
we obtained the $A_V$--distance dependence for the pulsar line of sight, which is shown 
in Fig. \ref{fig:Av_vs_dist_comp} by a red curve.
\begin{figure}
\begin{center}
\includegraphics[scale=0.3, clip]{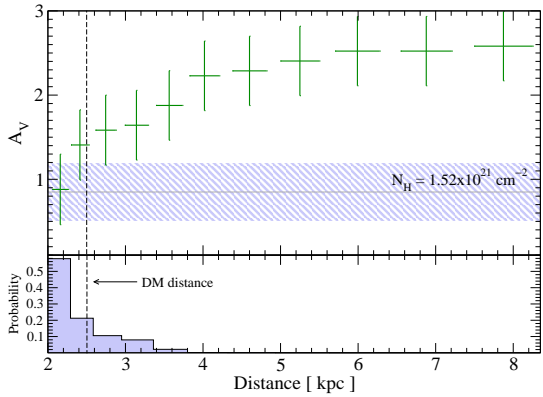}
\end{center}
 \caption{{\sl Top:} Comparison of the $A_V$--distance relation derived 
for the pulsar line of sight  (error-bars) and the $A_V$ range 
estimated  from  the absorbing column density $N_H$  obtained from the BB+PL 
spectral fit of the pulsar X-ray emission (hatched region). 
{\sl Bottom:} Probability distribution for the pulsar distance  derived from  
the intersection of the $A_V$--distance relation with  
the hatched region in the top panel. The vertical dashed line shows the distance 
of 2.5 kpc based on the DM measurements for the pulsar. 
 }
 \label{fig:Av_vs_dist}
\end{figure}

In the absence of significant extinction variations with  the Galactic longitude $l$,  
one can  enlarge  the region used to perform the extraction of the red clump stars,  to 
increase their number, and,   therefore, the  accuracy of the $A_V$--distance relation. 
However, this is impossible in our case.  Shifting the extraction region of the same size as above 
by $\pm$0.5 degrees  away of the pulsar along $l$,  we found a noticeable 
extinction variation with the longitude, 
demonstrated in Fig. \ref{fig:Av_vs_dist_comp}.  This is natural,  since the line of sight to the pulsar 
runs near and almost parallel to the  outer edge of the Scutum-Centaurus arm, 
where  the interstellar matter (ISM) density     
increases towards the arm centre.  As a result,  the extinction  is systematically 
higher for the region shifted towards  the arm centre (larger $l$), and is lower for the region  shifted 
away from the arm, as it  is seen in Fig. \ref{fig:Av_vs_dist_comp}. 
Therefore, owing to  the ISM 
density gradient  near the arm, any enlargement of the extraction 
region leads to systematic uncertainties, which 
do not allow us  to increase 
the accuracy of the $A_V$--distance relation.   
 
\citet{schleg98} provide $A_V$ = 5.9  for the entire Galactic extinction in this  direction, 
which  is about twice as high 
as  our   large distance limit, $A_V$ $ \la$ 3.  
However,  for the Galactic latitude $\la$ 5\fdg, as in our case,  
their  estimates can be unreliable. For instance, the 
Leiden/Argentine/Bonn (LAB)  Survey of Galactic HI \citep{kal05} 
indicates that the entire measured $N_{H}$ in the 
corresponding  direction  is       
as low as  6.6 $\times$ 10$^{21}$ cm$^{-2}$.  Using a standard     relation,  
$N_{H}$ = (1.79 $\pm$ 0.04) $\times$ 10$^{21} $ $A_{V}$  \citep{pred95}, 
this gives    $A_V$ = 3.7, which is considerably lower than that  of \citet{schleg98}, 
and  consistent with our value at a 2$\sigma$  significance.  
Owing to the vicinity of the Scutum-Centaurus arm, the entire hydrogen-column density varies  
by 30\%--40\%  with a position shift of a fraction of degree  around our line of sight,   
which increases towards the arm centre. This demonstrates that the current indirect estimates 
of the entire Galactic $A_V$  in this direction are  rather uncertain.     
 
At the same time, using the derived $A_V$--distance relation, we obtain distance 
estimates for  the main-sequence field stars  $A$ and $B$, which agree with the independent  
estimates carried out in Sect.  3.4 based on their positions in the colour-colour diagram.    
Selecting   $A_V$ $\approx$ 2.0 based on   the diagram (Fig.~\ref{fig:5}, bottom panel), 
the  $A_V$--distance relation gives us   
$\approx$ 4 kpc for these stars,  which is compatible with  a rough estimate of their distances $\sim$ 4.5 kpc,  
based on their spectral types estimated from the diagram.  
This test shows that our   $A_V$--distance relation can be reliably used at least for
distances  $\la$ 4--5 kpc.     
    
For  the DM distance to the pulsar  of 2.5 kpc, 
the relation provides $A_V$
$\approx$  1.5 $\pm$ 0.5  (Fig.~\ref{fig:Av_vs_dist}).  
At the same time, $N_H$ = 1.52$_{-0.56}^{+0.66}$ $\times$ 10$^{21}$ cm$^{-2}$ derived from 
the BB+PL X-ray spectral fit (see Table \ref{t:x-fit}) leads to a smaller value,  $A_V$ = 0.85 $\pm$ 0.33. 
Convolving  this  with the  $A_V$--distance relation,  
one can obtain a probability distribution for the distance to the pulsar 
corresponding to the  interval of  $A_V$  derived from  the fit (bottom panel of  Fig.~\ref{fig:Av_vs_dist}). 
It shows that the pulsar is likely to be closer to us by about 0.5 kpc than the DM  distance.   
The $N_H$ value   obtained 
from the NSA+PL model fit  suggests that $A_V$ $\approx$ 
1.9 $\pm$ 0.15, which is consistent with the DM distance, though  
the respective distance probability distribution is very wide and inconclusive. 
 
To summarise,  the   red clump stars allow us to constrain the $A_V$ value  
and distance for the pulsar   within  0.5--2.1 mag and   2--2.5 kpc ranges, respectively,  
depending on the model applied to fit  the X-ray spectrum of the pulsar.  
 \subsubsection{The optical-X-ray spectral energy distribution} 
\begin{figure}
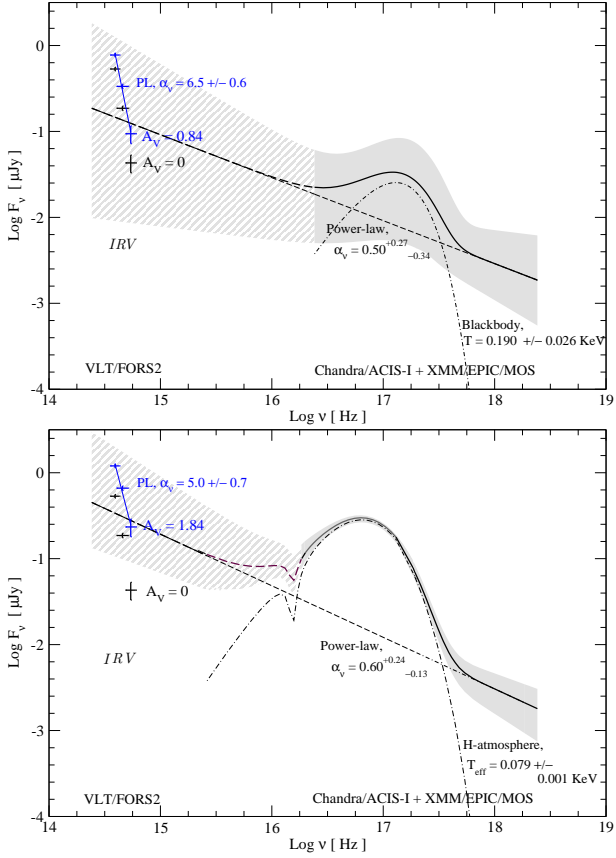

 \setlength{\unitlength}{1mm}
\resizebox{11.5cm}{!}{
\begin{picture}(50,50)(1,0)
\put (0,25) {\includegraphics[scale=0.15, clip]{mw_spectrum_1_reg.eps}}
\put (0,0) {\includegraphics[scale=0.15, clip]{mw_spectrum_3_reg.eps}}
 \end{picture}}
 \caption {Multi-wavelength unabsorbed spectra for  PSR J1357$-$6429, assuming the BB+PL  ({\sl top panel}) 
 and  NSA+PL  ({\sl bottom panel})  spectral models for  the explanation of its X-ray spectrum.  
 The best fits to the X-ray spectrum,
 including
 thermal (BB or NSA) and non-thermal (PL) components, 
 are shown by different lines. 
 The fits are extrapolated to the optical range.  
 The hatched regions are the fit and extrapolation uncertainties.    
 They are larger for the BB+PL  model  than for the NSA+PL   model, 
 because the NS radius to the distance ratio is fixed for the latter case, 
 which decreases the formal uncertainties in the fit.  The observed and dereddened optical fluxes 
 are shown by the black and blue cross-bars, respectively, and the blue lines are the best power 
 law fits to the dereddened optical SEDs. The respective  extinction values $A_V$,  
 spectral indices $\alpha_{\nu}$, 
 and NS temperatures $T$,  
 are shown in both plots.             
}
 \label{fig:mw1}
 \end{figure}

Possible unabsorbed optical-X-ray spectral energy distributions (SEDs) 
of the pulsar emission are presented 
in Fig.~\ref{fig:mw1}, where the optical fluxes are dereddened  in accordance  
with the results obtained in the previous subsection.  
As seen,  for any X-ray spectral model the intrinsic optical fluxes 
of the counterpart candidate are generally consistent,  
within the uncertainties,  
with the extrapolation of the X-ray spectrum 
towards longer wavelengths. However, 
the optical spectral slope is much steeper than the slope of the X-ray PL component, 
suggesting that there is a strong double-knee  spectral break between the optical and X-rays.  
Most  pulsars detected in the optical display a flatter spectrum in the optical than in X-rays,  
with a single break between these ranges. 
There are only two exceptions, 
PSR B0540$-$69.3 \citep{ser04}   and  \object{the Vela pulsar} \citep{Danila11}. 
The optical SED of \object{B0540$-$69.3}  also suggests 
that there is a double-knee spectral break, 
while its spectral index, $\alpha_{\nu}$ $\sim$ 1.1 
(assuming $F_{\nu} \propto \nu^{-\alpha_{\nu}}$), 
is  several times smaller than that for our optical candidate, 
$\alpha_{\nu}$ $\sim$ 4--6. 
For the Vela pulsar, the optical spectrum is also flat,  but in the near-infrared (near-IR) 
it has a steep flux increase 
with $\alpha_{\nu}$ $\sim$ 3,  
comparable to what we see for the J1357$-$6429 candidate in the optical.   
Near-IR flux excesses are also observed in the emission of 
two middle-aged pulsars \object{PSR B0656+14} and \object{Geminga} 
\citep{Danila11, shib06, durant11},  although their spectral slopes appear 
to be less steep than in the Vela case. If our candidate is 
the real optical counterpart of the pulsar J1357$-$6429, 
this would  represent the first example of an unusually steep 
spectrum  of  pulsar emission in the optical range.     

  \section{Discussion} \label{sec4}  
In the FORS2/FOV for our observations (Fig. \ref{fig:1}), 
the surface-area number density of observed point-like sources  
down to the limiting magnitude $R_{lim}$$\la$28 is   
$\sim$ 0.2 $object~arcsec^{-2}$. The probability of confusion  
of the pulsar counterpart with an unrelated point-like 
field optical object that accidentally  falls into the 90\% \textit{Chandra}  
positional uncertainty ellipse  of the pulsar with a radius of $\approx$0\farcs6  
(Fig.~\ref{fig:2}) is about  20\%. 
A similar estimate is obtained if we account for only  the immediate vicinity of the pulsar.   
This is not compelling evidence of a true identification, but 
from the multi-wavelength SED (Fig.~\ref{fig:mw1}) 
it is hard to expect  a  true   counterpart to be significantly 
brighter than our   candidate with  $R$$\approx$25.5.  
Pulsar optical fluxes typically do not significantly overshoot
the spectrum extrapolated from X-rays.    
The number density of observed sources in the  
brightness range of 25.5$\la$ $R$ $\la$28 expected for the counterpart  
is  much smaller, $\sim$0.0009 $object~arcsec^{-2}$ (cf. Fig.~\ref{fig:5}). 
The respective confusion probability is  $\sim$1\%  and 
becomes  considerably smaller, $\sim$0.02\%,  
if   we additionally constrain the colour $(R-I)$ $\la$ 1.4, 
as in the counterpart candidate case. 
This appears to be reasonably  compelling evidence of the association of the object $C$ with the pulsar. 
However,  the sample of  point-like objects in the considered magnitude range may be incomplete, 
and the confusion probability estimated in this way may be underestimated.   
It  indicates only that the real confusion probability is  somewhere between  $\sim$1\% and 20\%.    

A detailed analysis of the few field objects in the FOV, 
which fulfil the above magnitude constrants, shows that their 
locations in the colour-colour diagram are consistent 
with the main sequence distribution.  At the same time, 
another few objects located within the candidate colour uncertainties 
in the colour-colour diagram are brighter and lie within the stellar  
branch in the  colour-magnitude diagrams. 
Summarising  all  these results, we, therefore, cannot firmly reject       
the possibility that the detected  point-like optical  object  
is the optical counterpart
to the radio/X-ray/gamma-ray   pulsar J1357$-$6429. 
If it is indeed the counterpart, this implies 
at least two  prominent properties of the pulsar. 

   \begin{figure*}[t]
\includegraphics[scale=0.55, clip]{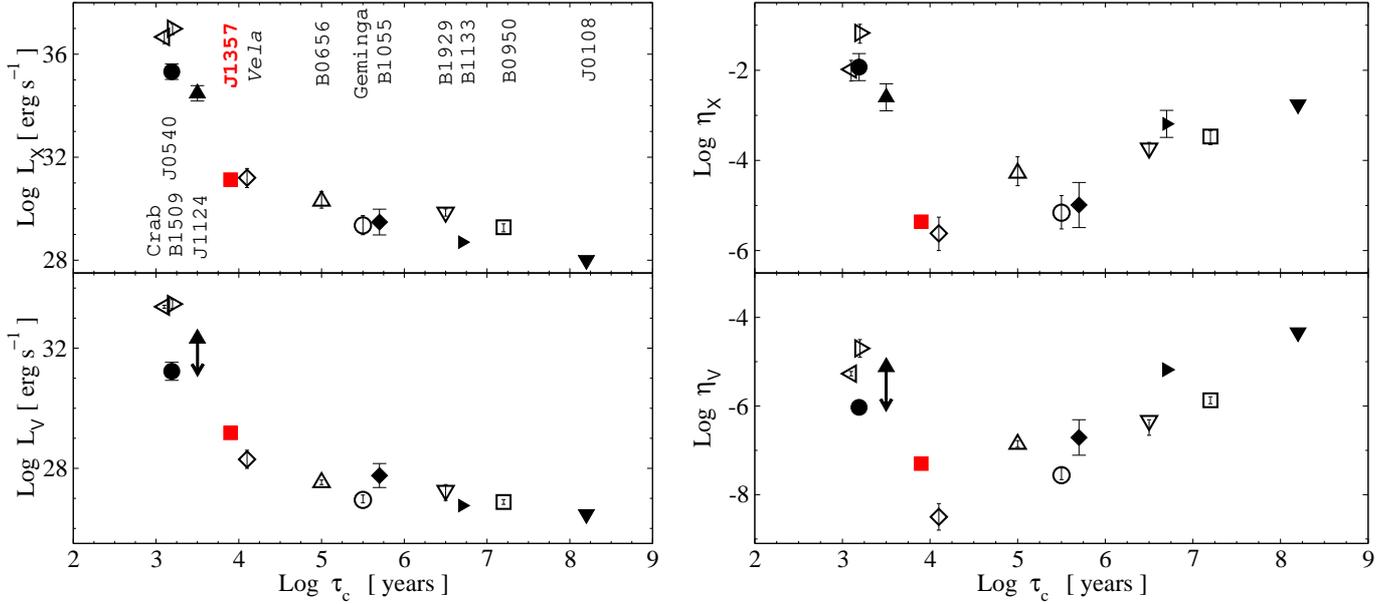}
 \caption {Comparison of  X-ray and $V$-band luminosities and efficiencies for pulsars 
of different characteristic age $ \tau$ 
detected in both spectral domains. Different pulsars are marked by different symbols. 
The tentative PSR J1357$-$642 optical 
luminosity and efficiency are shown by a filled square. As seen, 
both match the general dependences of the pulsar 
luminosity and efficiency on age, and display an efficiency minimum near the Vela age.    
}
 \label{fig:eff}
 \end{figure*}
     
Firstly,  a significant offset in the counterpart  candidate position 
from the  J1357$-$6429 radio coordinates   
measured  8.7 yr earlier, 1\farcs54 $\pm$ 0\farcs32 (Fig.~\ref{fig:2}),   
suggests that the pulsar has a high  proper motion (cf.  Sect. 3.1).       
At the most plausible distance range of 2--2.5 kpc,  this implies 
that the  transverse velocity of the pulsar is between    
1690 $\pm$ 350 km s$^{-1}$ and 2110 $\pm$ 440 km s$^{-1}$.
On the other hand, 
\citet{Abramowski11} obtained a two-to-three times smaller velocity,
650$d_{2.4}\tau_{7.3}^{-1}$ km s$^{-1}$, where $d_{2.4}$ is 
the distance to the pulsar in units of 2.4 kpc and
$\tau_{7.3}$ is the pulsar age in units of 7.3 kyr (characteristic age), 
based on a 7\amin~ offset (5$d_{2.4}$ pc) of the pulsar from 
the centre of the extended source HESS J1356$-$645, likely associated with an
"ancient" PWN and parent SNR, and assuming that the offset is due to the pulsar
proper motion.  However, the true ages of pulsars typically differ from 
characteristic ones by at least 40\% \citep[see, e.g.][]{brisken2003,thor2003}.
Therefore, we can reconcile the difference between the two estimates of the transverse velocity
assuming that true age of J1357$-$6429 is twice as small as the characteristic one.
We note that the position angle of the putative proper motion roughly coincides with 
the elongation of the X-ray plerion and HESS source images.
To our knowledge, the highest  pulsar velocity of 1080 $\pm$ 100 km s$^{-1}$,  
which has been firmly established  
by direct proper-motion and parallax measurements with the VLBA,   
belongs to \object{PSR B1508+55} \citep{chart2005}. 
The distance to that pulsar, 2.37 $\pm$ 0.20 kpc,  is comparable to that of J1357$-$6429, which suggests
that similar direct measurements   
are also possible for the latter. 
Another example of a high velocity pulsar is  B2224+65, 
which belongs to the Guitar  H$_{\alpha}$ bow-shock nebula.  
The radio interferometric measurements give a very precise proper motion for \object{PSR B2224+65} \citep{harr1993}, 
which corresponds to the velocity of about 1600  km s$^{-1}$, if we assume that the distance 
is in the range
1--2 kpc. Nevertheless, the distance has not been measured directly, by measurement of the parallax, and 
remains very uncertain \citep[see, e.g.][]{chart2004}.  
After the confirmation by radio interferometric observations,  J1357$-$6429  may become the 
fastest moving pulsar known. 
In this context, H$_{\alpha}$ imaging of the pulsar field would also be 
useful to search for a bow-shock nebula  
around the pulsar. 

It is  remarkable that   the direction of the  suggested proper motion 
for J1357$-$6429 (Fig.~\ref{fig:2}; Sect. 3.1) is consistent   
with the  NE extension of the  tail-like PWN structure detected 
in X-rays (Fig.~\ref{fig:4}). This is similar to what is seen 
for the Vela  and Crab pulsars, for which the 
proper motion direction virtually coincides  with their PWN jet-like structures;   this might imply
that the extended X-ray structure  of  J1357$-$6429 is  actually not a tail  but a pulsar jet. 
That is also supported  by the bend of the jet  
at  $\sim$~15\asec~ 
northward 
of the pulsar (\citet{Chang2011}, see Fig.~\ref{fig:4}),  
which is reminiscent of   the Vela-pulsar outer jet bend \citep{pavlov2003}.  
A  blob-like structure is visible 
in the middle of the J1357$-$6429 jet.  Variable blob-like structures are also typical of the Vela jet. 
This can be considered 
as indirect confirmation of the estimated  direction  of the  
putative 
J1357$-$6429  proper motion. 
       
Secondly, the putative counterpart   has an extremely red    spectrum,  with the spectral index 
$\alpha_{\nu}$ in the range of  4--6  (Fig.~\ref{fig:mw1}). This is atypical of pulsars and    
raises   doubts about the   pulsar nature of the candidate.  At an unrealistically large $A_V$  of $\sim$ 5.9,
comparable  to the entire Galactic extinction in this direction, the dereddened spectrum becomes flatter,    
$\alpha_{\nu}$ $\approx$ 2, but  the optical flux overshoots  the upper limit of the low energy 
extension of the pulsar X-ray PL emission component by an order of magnitude.  
This is also atypical of pulsars.  
As we have estimated from different points of view (Sect. 3.5.2), 
the real  $A_V$ for the pulsar is likely to be $\la$ 2. 
At the same time, the observed colour properties of the candidate are distinct from  those of field  stars, 
and any dereddening cannot place it on the main stellar sequence branch (Fig.~\ref{fig:5}), 
thus rejecting its association with an ordinary star. 
    
One of the possible explanations of the red spectrum is that the 
candidate is  a knot-like structure of the J1357$-$6429 PWN, 
or a combination of the emission from  the unresolved  pulsar and a nearby knot. 
Such  an optical knot, located  only 0\farcs65 of the pulsar, is known to exist 
in the Crab pulsar/PWN system \citep{hester95}. 
It is projected onto 
the pulsar's SE jet and has a power-law spectrum  with $\alpha_{\nu}$ $\sim$ 1.2, 
which is significantly steeper than the almost  flat optical spectrum of  the Crab pulsar  \citep{sol09}.     
Similar PWN knots  are likely  present near the Vela pulsar, where they have even  steeper spectra with 
$\alpha_{\nu}$~$\sim$~3 \citep{Danila11, shib03},  which is comparable to what we see for J1357$-$6429. 
This is a plausible interpretation, since the candidate position is  
shifted, albeit insignificantly, 
from the X-ray pulsar position by about 
0\farcs2 along the NE jet (Figs.~\ref{fig:2}, \ref{fig:4}). 

This steep  spectrum could also be produced by an  X-ray illuminated 
fall-back disk around the pulsar \citep{Perna2000}. 
The disks  are  presumably  formed   soon after the parent supernova explosions.  
However,  in our case this model  does not work  
because the pulsar X-ray luminosity is too low  
to  provide the optical flux density  at the observed  level  
after the reprocessing 
of X-rays to the optical in a hypothetical disk.     
In principle, an accreting fall-back disk with appropriate parameters can
fit the optical data. 
This model could be thoroughly examined if the pulsar nature of
the candidate were confirmed.

On the other hand, ignoring the spectral steepness and assuming that 
the candidate is the true pulsar optical counterpart, 
we can estimate the pulsar optical luminosity
in the $V$ band, 
$L_{V}$ = 1.5 $\times$ 10$^{29}$ ergs s$^{-1}$, 
and efficiency, $\eta_{V}$ = $L_{V}$/$\dot{E}$ 
= 4.8 $\times$ 10$^{-8}$, 
and compare them with those 
for other optical pulsars.  The comparison is shown in   Fig.~\ref{fig:eff},  
where we have also added the X-ray luminosities $L_{X}$ 
and efficiencies $\eta_{X}$ of the same pulsars.  One can see that our pulsar 
candidate 
does not look outstanding in these  integral 
dependencies. 
As expected, it lies close to 
the Vela pulsar, which is known to be 
under-luminous, or inefficient,   in the optical and X-rays 
and forms a pronounced minimum in the  $\eta_{V}$ and   
$\eta_{X}$ dependences with age \citep{zhar04,zhar06}.   
The ratio of $L_{X}$ to $L_{V}$ of about 
100 for our pulsar is 
also in the range of 100--1000 observed for other pulsars.  
All that supports the assumption that   J1357$-$6429 
is a Vela-like pulsar, as follows from its other properties 
(see Sect.~1), and is  indirect evidence that we 
detected the real optical counterpart of J1357$-$6429.    
       
An alternative interpretation could be a faint  extragalactic source,  such as an 
active galactic nucleus (AGN).    
The  spectral index of  about 2, as we have obtained above after dereddening with  
the entire Galactic extinction, is allowed for these  
objects  \citep{grupe10}.   However, no signs of  the putative AGN are seen in other spectral domains.  
Luminous X-ray emission is a primary signature of  AGN activity. 
The typical X-ray-to-optical flux ratios, $f_X$/$f_r$, 
of AGNs  obey a condition $\log(f_X/f_r)$ $\ga$ $-$0.5, or $\ga$ $+$0.1 
for blazars \citep{green04}.    Using the dereddened source magnitudes $R$ $\approx$ 20.68 and   
$I$ $\approx$ 20.32, we converted them into the SDSS $r$ band magnitude of $\approx$ 20.85. 
On the basis of the above $ f_X/f_r$ constraints  and the formula from  \citet{green04},   
$\log(f_X) =\log(f_X/f_r)  -5.67-0.4r$, we 
estimated a lower $f_X$ limit to the 0.5--2 keV range from possible AGN,  
$f_X$ $\sim$  (0.3--1.2) $\times$ 10$^{-14}$ ergs  s$^{-1}$ cm$^{-2}$, which is 
at least a factor of five lower than 
the unabsorbed X-ray flux detected from the pulsar/PWN system in the same range, $f_X$ $\sim$ 
(6--14) $\times$ 10$^{-14}$ ergs s$^{-1}$ cm$^{-2}$,  depending on  the model accepted for 
the pulsar thermal emission component.    
In the limiting case,  the putative 
AGN would  noticeably  contaminate the pulsar/PWN  X-ray flux by 20\%,   
but we do not resolve any background 
point-like X-ray object at the pulsar position  except for the pulsar.  
This means that a putative AGN or a galaxy
is  fainter in X-rays than the pulsar  by an order of magnitude,  which  
cannot be completely excluded by the above estimates and  current data. 
AGNs and galaxies occupy a   
region that is very distinct from that of stars in  the infrared colour-colour diagram
\citep{kim11}, and  multi-band infrared 
observations  can help us to clear the real nature of the object.   

To summarise,  we have detected  a candidate  optical counterpart of the pulsar J1357$-$6429. 
It appears to have an extremely high  transverse velocity  
and an unusually  steep optical spectrum. 
Radio interferometric observations are necessary to confirm the high proper motion of the pulsar and 
measure its parallax and distance.  High resolution imaging provided by the \textit{HST} 
or ground-based telescopes with AO systems would be useful to resolve 
a possible PWN knot near  the pulsar position, 
which would explain   
the steepness of the spectrum.  
It can also help us to measure the candidate proper motion. 
If the results of this measurement  
coincide with those of
forthcoming radio measurements of the pulsar proper motion, this
would be a firm confirmation of the pulsar nature of the optical 
object. The H$_{\alpha}$  imaging would  allow us to search for a bow-shock 
nebula around such a high velocity pulsar and to constrain independently 
the parameters of the pulsar wind. 
Deep infrared imaging of the  pulsar field is crucial to  
extend the   SED of the candidate towards longer wavelengths and understand 
whether it is really associated with the pulsar 
or it is a distant background galaxy or AGN.

\begin{acknowledgements}
We are grateful to an anonymous referee for useful comments
improving the paper.
The work was partially supported by the Russian Foundation for Basic 
Research (grants 11-02-00253 and 11-02-12082),
Rosnauka (Grant NSh 3769.2010.2), and 
the Ministry of  Education and Science of the Russian Federation (Contract No. 11.G34.31.0001).
SZ acknowledges support from CONACYT 151858  project.
The work by GP was partly supported by NASA grant NNX09AC84G.
REM acknowledges support by  the BASAL
Centro de Astrof\'isica y Tecnologias Afines (CATA) PFB--06/2007.
We used the USNOFS Image and Catalogue Archive operated by the United States Naval Observatory, 
Flagstaff Station (http://www.nofs.navy.mil/data/fchpix/).  
The Munich Image Data Analysis System ({\tt MIDAS}) is developed and
maintained by the European Southern Observatory. 
Image Reduction and Analysis Facility     
({\tt IRAF}) is distributed by the National Optical Astronomy Observatories,
which are operated by the Association of Universities for Research
in Astronomy, Inc., under cooperative agreement with the National
Science Foundation.
\end{acknowledgements}

\end{document}